\documentclass[a4paper,11pt,dvips]{article}
\usepackage{latexsym}
\usepackage{amsmath}
\usepackage{a4wide}
\usepackage{graphicx, subfigure}
\usepackage{cite}

\newcommand{\pslash}{p \hspace{-4pt} / }

\begin{document}

\title{\vspace{-2cm}
\begin{flushright}
\small
TTP11-12 \\
SFB/CPP-11-21\\
MZ-TH/11-07
\end{flushright}
\vspace{2cm}
\Large\bf  Complete next-to-leading order gluino contributions
    to $b\to s \gamma$ and $b \to s g$
  }
\author{Christoph Greub$^a$, Tobias Hurth$^b$, Volker Pilipp$^a$, Christof Sch\"upbach$^a$,\\
and Matthias Steinhauser$^c$\\[4mm]
{\small $^a\,$Albert Einstein Center for Fundamental Physics,}\\[-1.0mm]
{\small Institute for Theoretical Physics, University of Bern,}\\[-1.0mm]
{\small 3012 Bern, Switzerland}\\[1mm]
{\small $^b\,$Institute for Physics (THEP) ,}\\[-1.0mm]
{\small  Johannes Gutenberg University,}\\[-1.0mm]
{\small 55099 Mainz, Germany}\\[1mm]
{\small $^c\,$Institut f\"ur Theoretische Teilchenphysik,}\\[-1.0mm]
{\small   Karlsruhe Institute of Technology (KIT)}\\[-1.0mm]
{\small 76128 Karlsruhe, Germany}
}
\date{}
\maketitle
\begin{abstract}
We present the first complete order $\alpha_s$ corrections 
to the Wilson coefficients (at the high scale) of the various versions of
magnetic and chromomagnetic operators which are induced by a
squark-gluino exchange. For this matching calculation, we work out the on-shell amplitudes  
$b\to s\gamma$ and $b \to s g$, both in the full
and in the effective theory up to order $\alpha_s^2$.
The most difficult part of the calculation is the evaluation of the 
two-loop diagrams in the full theory; these can be split into two
classes: a) diagrams with one gluino and a virtual gluon; b)
diagrams with two gluinos or with one gluino and a four-squark
vertex. 
Accordingly, the Wilson coefficients can be split into a part~a) and a
part~b).
While part~b) of the Wilson coefficients is presented in this paper
for the first time, part~a) was given in \cite{Bobeth:1999ww}.
We checked their results for the coefficients of the magnetic operators and
found perfect agreement. 
Moreover, we work out the renormalization procedure in great detail.

Our results for the complete next-to-leading order  Wilson coefficients are fully
analytic, but
far too long to be printed. We therefore publish them in the form
of a C++ program. They constitute a crucial  building block for the phenomenological
next-to-leading logarithmic analysis of the branching ratio $\bar B \rightarrow X_s \gamma$ in a
supersymmetric model beyond minimal flavor violation.
    
\end{abstract}
\thispagestyle{empty}

\vfill
\newpage
\thispagestyle{empty}
\tableofcontents
\setcounter{page}{1}
\section{Introduction}

Among the rare $B$ meson decays those induced by 
radiative and electroweak penguin diagrams are 
of particular interest.
Such   flavor changing neutral current (FCNC) processes  
offer high sensitivity to new physics (NP) through potential new
degrees of freedom beyond the Standard Model (SM).
  Additional contributions to the
decay rate, in which SM particles in the loops are replaced by new particles such as
the supersymmetric charginos or gluinos are not suppressed by the loop 
factor $\alpha/4\pi$ relative to the SM contribution. Thus,  FCNC
decays provide information about the SM and its extensions via virtual
effects to scales presently not accessible otherwise. This approach is complementary to the
direct production of new particles at collider experiments (for reviews
see~\cite{Hurth:2010tk,Hurth:2003vb}).

At the end of the first generation of the $B$ factories at KEK (Belle
experiment at the KEKB $e^+e^-$ collider)~\cite{Belle} and at SLAC (BaBar
experiment at the PEP-II $e^+e^-$ collider)~\cite{Babar},
all present measurements in flavor physics including those from the
Tevatron $B$ physics programs (CDF~\cite{TevatronB1} and
D0~\cite{TevatronB2} experiments) have not observed any unambiguous sign
of new physics, 
in particular  no ${\cal O}(1)$ NP effects in any FCNC 
process~\cite{Antonelli:2009ws, Buchalla:2008jp}.  This implies
the famous flavor problem, namely why FCNC processes are suppressed. 
It has to be solved in any viable new physics model.
It is well-known that the hypothesis of minimal flavor 
violation (MFV)~\cite{D'Ambrosio:2002ex,Hurth:2008jc}, i.e.\ that the NP model
has no flavor structures beyond the Yukawa couplings, solves the problem formally.
However,   new flavor structures beyond the Yukawa couplings are still compatible with the present data~\cite{Hurth:2009ke}   because the flavor sector has been tested only 
at the $10\%$  level, especially in the $b\to s$  transitions.

Among the  penguin modes, the inclusive decay $\bar B \rightarrow X_s \gamma$  is the 
most important one, because it is theoretically 
well understood and at the same time it  has been
 measured extensively at the $B$ factories.
While non-perturbative  corrections to this decay mode are subleading and were
recently estimated to be well below $10\%$~\cite{Benzke:2010js},  
perturbative QCD corrections are the most important corrections.
Within a global effort,  a  perturbative QCD calculation to the
next-to-next-to-leading-logarithmic (NNLL) level
has been
performed and has led to the first NNLL prediction of the $\bar B \to X_s  \gamma$ branching 
fraction~\cite{Misiak:2006zs,Misiak:2006ab} which also includes the nonperturbative
contributions. Using the photon energy cut $E_0=1.6$ GeV,  the branching
ratio reads 
\begin{equation}\label{final1}
{\cal B}(\bar B \to X_s \gamma)_{\rm NNLL} =  (3.15 \pm 0.23) \times 10^{-4}.
\end{equation}
The combined experimental data according to the Heavy Flavor  Averaging Group
(HFAG)~\cite{hfag} leads to 
\begin{equation}
 {\cal B}(\bar B \rightarrow X_s \gamma) = (3.55 \pm 0.24 \pm 0.09) \times 10^{-4}
 \,,
\end{equation}
where the first error is combined statistical and systematic, and the second
is due to the extrapolation in the photon energy.  
Thus, the SM prediction and the experimental average are consistent at the $1.2 \sigma$ level.
This implies very stringent constraints on NP  models like the
bound on the charged Higgs mass in the two Higgs-doublet model
(II)~\cite{Ciuchini:1997xe,Borzumati:1998tg} ($M_{H^+} > 295{\rm GeV}$
at $95\%$ CL)~\cite{Misiak:2006zs} or the bound on the inverse
compactification radius of the minimal universal extra dimension model
(mACD) ($1/R > 600 {\rm GeV}$ at $95\%$ CL)~\cite{Haisch:2007vb}.  In
both cases the bounds are much stronger than the ones derived from other
measurements.  Constraints within various supersymmetric extensions are
analyzed in \cite{Bertolini:1990if, Degrassi:2000qf,
  Carena:2000uj, Borzumati:1999qt, Besmer:2001cj,
  Ciuchini:2002uv, Ciuchini:2003rg, Okumura:2003hy, Degrassi:2006eh, Ciuchini:2007ha,
  Altmannshofer:2008vr, Crivellin:2009ar} 
 (for overviews
see~\cite{Hurth:2003vb,Altmannshofer:2009ne}).  Bounds on the Little
Higgs Model with T-parity have also been presented~\cite{Blanke:2009am}.
Finally, model-independent analyses in the effective field theory
approach without~\cite{Ali:2002jg} and with the assumption of minimal
flavor violation~\cite{ D'Ambrosio:2002ex, Hurth:2008jc} also show the
strong constraining power of the $\bar B \rightarrow  X_s \gamma$ branching
fraction.

While within the perturbative SM calculation NNLL precision has been
achieved~\cite{Misiak:2006zs,Misiak:2006ab}, also within
supersymmetric theories higher order calculations have been pushed forward in
recent years.
In particular, the complete NLL QCD calculation of the MSSM with the additional
assumption of minimal flavor violation was presented in \cite{Degrassi:2006eh}
including a published computer code~\cite{Degrassi:2007kj}.
Beyond minimal flavor violation, the most important role is played by the
non-diagonal gluino-quark-squark vertex due to the large strong coupling which
comes with this vertex. 
This flavor non-diagonal vertex is induced by squark-mixing to the extent as it
is misaligned with quark mixing.  It represents  a  new flavor structure
beyond the SM Yukawa couplings. 
A complete LL analysis of the corresponding gluino contribution to the
inclusive decay rate was  
presented in \cite{Borzumati:1999qt}. 
Some ``beyond LL effects'' 
were estimated~\cite{Okumura:2003hy}, but a general NLL analysis of the gluino
contribution is still missing.  

Besides these NLL contributions due to the gluino vertex, 
there are of course more NLL corrections with non-minimal flavor violation;
they involve electroweak (gaugino and higgsino) vertices.
However, such contributions are in general suppressed compared to the 
corrections computed here.
There are two types of such contributions at the NLL level:  First, there are
electroweak corrections to the non-minimal LL gluino contribution (in which 
the electroweak vertex is flavor-diagonal or MFV-like) which are naturally
suppressed due to the smaller coupling constants and due to the CKM hierarchy. 
Second, there is also non-minimal flavor violation via squark-mixing 
in the electroweak vertices possible. 
But such contributions are already
 suppressed at the LL level compared to the gluino contribution  
due to the smaller  coupling constant, apart from the chargino contributions in specific  parts of the parameter space 
in which for example the trilinear coupling $A^u_{23}$ is  very large. 
These features  do not change of course when gluon- and gluino-induced  NLL corrections are added 
to such  LL contributions.\footnote{Still, the leading chirally enhanced corrections can be
easily calculated by 
inserting the effective Feynman rules of \cite{Crivellin:2011jt}
into the results of \cite{Besmer:2001cj}.}
Summing up, the complete NLL corrections induced by the gluino vertex represent the dominant 
contribution beyond MFV at this order in most parts of the MSSM parameter space. 
They  are complementary to the MFV contributions at the NLL level which are given  in \cite{Degrassi:2006eh}.

In the present paper we work out for the first time the complete
corrections of order $\alpha_s$
to the Wilson coefficients (at the matching scale $\mu_W$) of the various versions of
magnetic and chromomagnetic operators which are induced by a
squark-gluino loop. In our procedure all the appearing heavy particles (which are the gluino, the
squarks and the top quark) are simultaneously integrated out at the high
scale, which we call $\mu_W$.
For this matching calculation, we work out the on-shell amplitudes  
$b\to s\gamma$ and $b \to s g$, both in the full
and in the effective theory up to order $\alpha_s^2$.
The most difficult part of the calculation is the evaluation of the 
two-loop diagrams in the full theory; these can be split into two
classes: a)  diagrams with one gluino and a virtual gluon; b)
diagrams with two gluinos or with one gluino and a squark-loop.
Accordingly, the Wilson coefficients can be split into a part~a) and a
part~b).
While part~b) of the Wilson coefficients is given in the present work 
for the first time, part a) was published by Bobeth et al.   \cite{Bobeth:1999ww}.
We explicitly checked their results for the Wilson coefficients of the
magnetic operators, which were obtained using an off-shell matching procedure,
and found perfect agreement.

Moreover, we work out all steps of the NLO renormalization procedure in great detail;
to our knowledge, the complete set of renormalization constants, which we give in
Appendices~\ref{sec:quarkrenormalization} to \ref{renormg}, 
have not been presented before.
We also discuss how the recently discussed chirally-enhanced terms \cite{Crivellin:2009ar}  
enter our scheme.
Our results for the complete NLO  Wilson coefficients are fully
analytic, but
far too long to be printed. We therefore publish them in the form
of a C++ program. 

The remainder of the paper is organized as follows: In the next section we
present details on the framework used for our calculation. In particular, we
list all relevant operators contributing to $\bar{B}\to X_s\gamma$.
In Section~\ref{sec:magnetic}, we describe in detail the calculation of the
Wilson coefficients of the magnetic operators which is extended to the
chromomagnetic case in Section~\ref{sec:chromo}. All our results are
implemented in a computer code which is described in
Section~\ref{sec:results} and we present our conclusions in
  Section~\ref{sec:concl}. 
In Appendix~\ref{sec:LOresult}, we present explicit analytic expressions for
the LO result and 
Appendices~\ref{sec:quarkrenormalization}
to~\ref{renormg} provide details on the NLO renormalization procedure.

\section{Preliminaries}

\subsection{Squark-quark-gluino interaction}

The minimal supersymmetric SM (MSSM) allows for generic new sources of flavor violation beyond the
Yukawa structure in the SM, thus, beyond the minimal flavor violation hypothesis. Besides the usual quark mixing also the squarks induce flavor mixing
due to a possible misalignment of quarks and squarks in flavor space.

More explicitly, in the MSSM without soft SUSY-breaking, 
flavor-violation is induced by the superpotential terms
containing the Yukawa couplings of
the matter superfields to the Higgs superfields
\begin{equation}
{\cal L}^{\rm Yukawa} =   -\bar{\cal D}^o_{i} (h_d)_{ij} {\cal Q}^o_j {\cal H}_d + 
\bar{\cal U}^o_{i} (h_u)_{ij} {\cal Q}^o_j {\cal H}_u 
-\mu {\cal H}_d {\cal H}_u.
\label{superpotential}
\end{equation}
The index ``$o$'' indicates that the current eigenstate basis is used. 
Following the notation of \cite{Martin:1997ns,Allanach:2008qq}, 
we define   $\tilde{u}^{o*}_R$  as   the scalar component 
and $u_R^{o \dagger}$ as the fermionic component of the superfield $\bar{U}^o$ (and for $\bar{D}^o$
analogously); the scalar and fermionic components of the superfield ${\cal Q}^o$
are  defined as $\tilde{Q}^o = (\tilde{u}_L^o  \tilde{d}_L^o)$ and $Q=(u_L^o   d_L^o)$, respectively.
$\mathcal{H}_{u,d}$ are the Higgs superfields. The matrices $h_d$
and $h_u$ act on flavor space. Diagonalizing these matrices via
bi-unitary transformations defines the super-CKM basis.

In addition, there  are {\it soft}  SUSY-breaking terms leading to flavor
transitions of the quarks. 
When restricted to the terms relevant for squark masses
and quark-flavor transitions, the soft part of the Lagrangian can be 
expressed in terms of the scalar fields in the current eigenstate basis:
\begin{equation}
\begin{split}
{\cal L}^{\rm soft}  =&  -\tilde{Q}^{o \dagger}_i m_{\tilde{Q},ij} \tilde{Q}^o_j
-\tilde{u}_{Ri}^{o} m_{\tilde{u},ij}^2 \tilde{u}^{o *}_{Rj}
-\tilde{d}_{Ri}^{o} m_{\tilde{d},ij}^2 \tilde{d}^{o *}_{Rj}\\
&- \left(A_{ij}^d H_d \tilde{Q}^o_i\tilde{d}_{Rj}^{o *}
 + A_{ij}^u H_u \tilde{Q}^o_i\tilde{u}_{Rj}^{o *}  +\mbox{h.c.}\right),
\label{softsusybreaking}
\end{split}
\end{equation}
where $\tilde{u}^o$, $\tilde{d}^o$ and $\tilde{Q}^o$ are the  squark fields
  and $m_{\tilde{u}}$, $m_{\tilde{d}}$ and $m_{\tilde{Q}}$ the
  corresponding hermitian mass matrices. $H_u$ and $H_d$ are Higgs doublets and
  $A_u$ and $A_d$ are $3\times3$ matrices in flavor space.
For the  latter  
no proportionality to the Yukawa couplings is assumed; they are left
completely general.

After the diagonalization of the Yukawa couplings
in (\ref{superpotential}),
i.e. in the super-CKM basis, 
the squark mass matrices are of the form ($q=u$ or $q=d$)
\begin{equation}
{M}_{q}^2 =  
\left( \begin{array}{cc}
  m^2_{q,LL}   & m^2_{q,LR} \\
  m^2_{q,RL}  &  m^2_{q,RR}                 
 \end{array} \right) +
\label{squarku}
  \left( \begin{array}{cc}
  F_{q,LL} +D_{q,LL} &  F_{q,LR} \\
 F_{q,RL} & F_{q,RR} +D_{q,RR}                
 \end{array} \right),
\end{equation} 
where the  $F$-terms from the superpotential and the $D$-terms from
the gauge sector are diagonal $3 \times 3$ submatrices of the 
$6 \times 6$ mass matrices ${M}^2_q$. This is in general not true 
for the additional terms, originating from  the soft 
part of the Lagrangian, given in (\ref{softsusybreaking})  (for more details 
see Chapter III of \cite{Borzumati:1998tg}). 
Thus, the  (hermitian) squark mass terms  
\begin{equation}
(\tilde{u}^{\dagger}_L,\tilde{u}^{\dagger}_R) M^2_u 
 \genfrac(){0pt}{}{\tilde{u}_L}{\tilde{u}_R}			
\quad \text{and} \quad
(\tilde{d}^{\dagger}_L,\tilde{d}^{\dagger}_R) M^2_d 
  \genfrac(){0pt}{}{\tilde{d}_L}{\tilde{d}_R},
\end{equation} 
where the fields $\tilde{u}_{L,R}$ and $\tilde{d}_{L,R}$ now refer to 
the super-CKM basis, are not diagonal in general. Only when performing
an additional appropriate unitary field transformation defined
through
\begin{equation}
    \genfrac(){0pt}{}{\tilde{q}_L}{\tilde{q}_R} = 
    \left(\genfrac{}{}{0pt}{}{\Gamma_{QL}^{\dagger}}{\Gamma_{QR}^{\dagger}}\right)
    \tilde{q}
    \,,
\end{equation}
with $6\times 3$ matrices $\Gamma_{(U,D)(L,R)}$, one obtains 
a diagonal matrix 
\begin{equation}
  (\Gamma_{QL},\Gamma_{QR}) M^2_q  
  \left(\genfrac{}{}{0pt}{}{\Gamma_{QL}^{\dagger}}{\Gamma_{QR}^{\dagger}}\right)\,,
  \label{squark_rotation_1}
\end{equation}
and the squark mass terms have the form
\begin{equation}
  \tilde{q}^{\dagger} 
  (\Gamma_{QL},\Gamma_{QR}) M^2_q  
  \left(\genfrac{}{}{0pt}{}{\Gamma_{QL}^{\dagger}}{\Gamma_{QR}^{\dagger}}\right)
  \tilde{q}\,.
  \label{squark_rotation}
\end{equation} 
$\tilde{q}$ represents  the six squark fields  
in the mass-eigenstate basis. Note
that in (\ref{squark_rotation_1}) and (\ref{squark_rotation}) one
either has $(q,Q)=(u,U)$ or $(q,Q)=(d,D)$.
The $\Gamma$ matrices satisfy the following unitarity relations:
\begin{equation}
\sum_{i=1}^3 (\Gamma_{QL}^{ki} \Gamma_{QL}^{hi*} + \Gamma_{QR}^{ki}
\Gamma_{QR}^{hi} ) =  
\delta^{kh},   \quad \quad \sum_{k=1}^6 \Gamma_{QX}^{ki} \Gamma_{QY}^{kj*} = \delta^{ij} \delta_{XY},
\label{unitarityrelations}
\end{equation}
where $Q=U,D$ and $X,Y=L,R$.

The squark-quark-gluino coupling is flavor-diagonal in the super-CKM
basis; thus, the last rotation of the squark fields of (\ref{squark_rotation})   gives rise to the following 
squark-quark-gluino vertices,
which are non-diagonal in flavor-space
\begin{align}
\raisebox{-0.8cm}{\resizebox{4cm}{!}
{\includegraphics{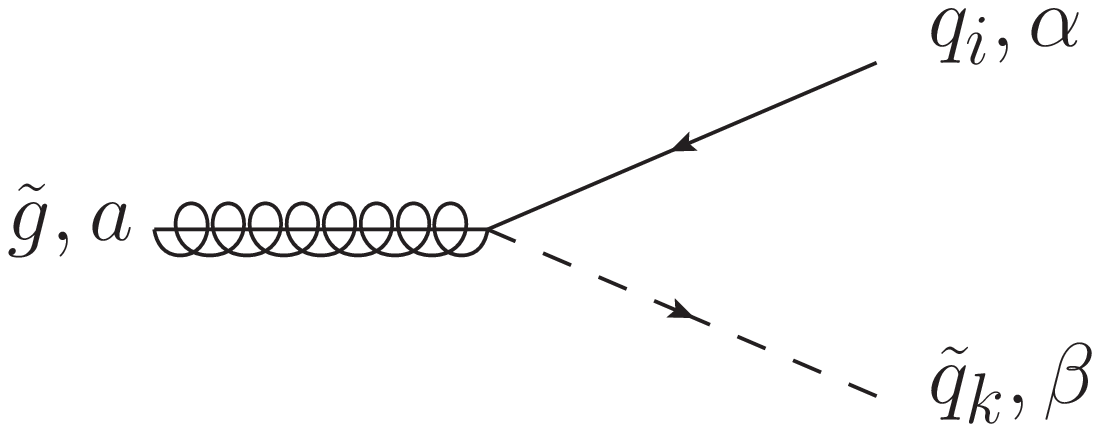}}}
\quad\quad &
-ig_s\sqrt{2}T^a_{\beta\alpha}
\left(\Gamma_{QL}^{ki} P_L - \Gamma_{QR}^{ki} P_R \right),
\label{sqg_vertex1}\\
\raisebox{-0.8cm}{\resizebox{4cm}{!}
{\includegraphics{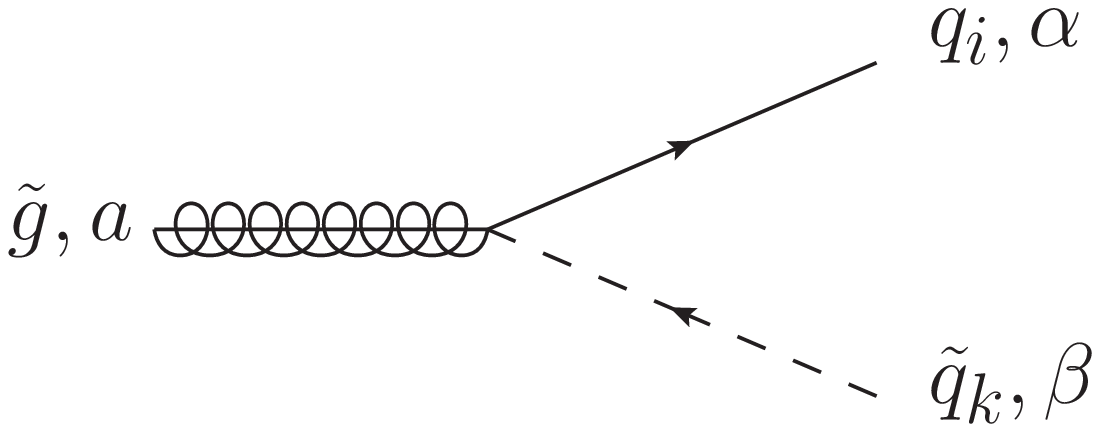}}}
\quad\quad &
-ig_s\sqrt{2}T^a_{\alpha\beta}
\left(\Gamma_{QL}^{ ki*} P_R - \Gamma_{QR}^{ki*} P_L \right),
\label{sqg_vertex2}
\end{align}Further Feynman rules may be found in 
\cite{Rosiek:1989rs,Rosiek:1995kg,Denner:1992vza}.

At this point, a comment on the super-CKM basis is appropriate. While in
\cite{Okumura:2003hy} and \cite{Degrassi:2006eh} a definition of the super-CKM basis is used in which the physical quark
 mass matrices 
are diagonal, we will apply a minimal renormalization scheme to
 the squark mixing matrices $\Gamma$ appearing in (\ref{sqg_vertex1})
 and (\ref{sqg_vertex2}). 
This corresponds to a  definition of the super-CKM basis in
which the Yukawa couplings 
remain diagonal also at the loop level \cite{Crivellin:2008mq,Hofer:2009xb,Crivellin:2009ar}. 

In the following calculation, we will express our results 
for the Wilson coefficients in terms of the diagonal squark masses
$m_{\tilde{q}_k}$ and the $\Gamma$ matrices, renormalized in the $\overline{\mbox{DR}}$ scheme.

\subsection{Operator basis}

In supersymmetric models, various combinations of the gluino--quark--squark vertex lead to
$|\Delta(B)|=|\Delta(S)|=1$ magnetic and chromomagnetic
operators (of ${\cal O}_7$-type, ${\cal O}_8$-type) with an explicit 
factor $\alpha_s$, and to four-quark operators, with a factor 
$\alpha_s^2$ when integrating out squarks and
gluinos~\cite{Borzumati:1999qt}. These squark/gluino induced operators
define the effective Hamiltonian
$\mathcal{H}_{\text{eff}}^{\tilde{g}}$, which is given in (\ref{HEffGluino}) below.
 
Using the standard effective field theory approach, the aim 
is to resum the following terms:
\begin{itemize}
\item[] \quad LO:
 $\quad \quad \alpha_s \, (\alpha_s L)^N$  
\item[] \quad NLO: 
 $\quad \alpha_s \, \alpha_s (\alpha_s L)^N, \quad \quad (N=0,1,...),$
\end{itemize}
respectively at the leading and next-to-leading order where 
$L$ denotes large logarithms of the ratio $m_b^2/M^2_{\rm{susy}}$. 

As discussed in \cite{Borzumati:1999qt},     
${\cal H}_{eff}^{\tilde{g}}$ is unambiguous, but it is a matter of
convention whether the $\alpha_s$ factors, peculiar to the gluino 
exchange, should be put into the
definition of operators or into the Wilson coefficients. 
It is  convenient to distribute the factors of $\alpha_s$ between operators
and Wilson coefficients in such a way that the 
matching calculation and the evolution down to
the low scale $\mu_b$ of the Wilson coefficients are organized exactly
in the same way as in the SM such that the anomalous-dimension matrix indeed
has the canonical expansion in $\alpha_s$ and starts with a term
proportional to $\alpha_s^1$.  

The effective Hamiltonian $\mathcal{H}_{\text{eff}}^{\tilde{g}}$ 
is of the form
\begin{equation}
\mathcal{H}_{\text{eff}}^{\tilde{g}}=
\sum_i C_{i\tilde{g}}(\mu) \mathcal{O}_{i\tilde{g}}(\mu)+
\sum_i \sum_q C_{i\tilde{g}}^q(\mu) \mathcal{O}_{i\tilde{g}}^q(\mu).
\label{HEffGluino}
\end{equation}
In the second term, we sum over all light quark flavors
$q=u,d,c,s,b$. 

At dimension five the following two-quark operators contribute
\begin{equation}         
\begin{array}{llll}  
{\cal O}_{7\tilde{g},\tilde{g}}         \,= &\!
  e \,g_s^2(\mu) \,
 (\bar{s} \sigma^{\mu\nu} P_R b) \, F_{\mu\nu}\,,    
                                        &  \quad 
{\cal O}_{7\tilde{g},\tilde{g}}^\prime  \,= &\!
  e \,g_s^2(\mu) \,
 (\bar{s} \sigma^{\mu\nu} P_L b) \, F_{\mu\nu}\,,     \\
                                        &             \\[-1.3ex]           
{\cal O}_{8\tilde{g},\tilde{g}}         \,= &\!
 g_s(\mu) \,g_s^2(\mu) \,
 (\bar{s} \sigma^{\mu\nu} T^a P_R b)
     \, G^a_{\mu\nu}\,, 
                                        &  \quad 
{\cal O}_{8\tilde{g},\tilde{g}}^\prime  \,= &\!
 g_s(\mu) \,g_s^2(\mu) \,
 (\bar{s} \sigma^{\mu\nu} T^a P_L b)
     \, G^a_{\mu\nu}\, ,
\label{gmagnopfive}                                     
\end{array} 
\end{equation}
with $P_{L}=\frac{1}{2}(1-\gamma_5)$ and $P_{R}=\frac{1}{2}(1+\gamma_5)$.
In the SUSY contributions corresponding to these operators, the chirality-violating parameter is the 
gluino mass $m_{\tilde{g}}$ which is included in the corresponding Wilson coefficients:

At dimension six the two-quark operators 
with chirality violation coming 
from the $b$- or $c$-quark mass 
read:
\begin{equation}
\begin{array}{llll}
{\cal O}_{7b,\tilde{g}}                 = &\!\!\!\!
   e \,g_s^2(\mu) \,{\overline m}_b(\mu) \,
 (\bar{s} \sigma^{\mu\nu} P_R b) \, F_{\mu\nu}\,,   
                                        &  
{\cal O}_{7b,\tilde{g}}^{\prime}        = &\!\!\!\!
   e \,g_s^2(\mu) \,{\overline m}_b(\mu) \,
 (\bar{s} \sigma^{\mu\nu} P_L b) \, F_{\mu\nu}\,,     \\[2.0ex]    
{\cal O}_{8b,\tilde{g}}                 = &\!\!\!\!
 g_s(\mu) \,g_s^2(\mu) \,{\overline m}_b(\mu) \,
 (\bar{s} \sigma^{\mu\nu} T^a P_R b)
     \, G^a_{\mu\nu}\,,                               
                                        &  
{\cal O}_{8b,\tilde{g}}^\prime          = &\!\!\!\!
 g_s(\mu) \,g_s^2(\mu) \,{\overline m}_b(\mu) \,
 (\bar{s} \sigma^{\mu\nu} T^a P_L b)
     \, G^a_{\mu\nu}\,,     \\[2.0ex]                          
{\cal O}_{7c,\tilde{g}}                 = &\!\!\!\!
   e \,g_s^2(\mu) \,{\overline m}_c(\mu) \,
 (\bar{s} \sigma^{\mu\nu} P_R b) \, F_{\mu\nu}\,,   
                                        &  
{\cal O}_{7c,\tilde{g}}^{\prime}       = &\!\!\!\!
   e \,g_s^2(\mu) \,{\overline m}_c(\mu) \,
 (\bar{s} \sigma^{\mu\nu} P_L b) \, F_{\mu\nu}\,,     \\[2.0ex]    
{\cal O}_{8c,\tilde{g}}                 = &\!\!\!\!
 g_s(\mu) \,g_s^2(\mu) \,{\overline m}_c(\mu) \,
 (\bar{s} \sigma^{\mu\nu} T^a P_R b)
     \, G^a_{\mu\nu}\,,                               
                                        &  
{\cal O}_{8c,\tilde{g}}^\prime          = &\!\!\!\!
 g_s(\mu) \,g_s^2(\mu) \,{\overline m}_c(\mu) \,
 (\bar{s} \sigma^{\mu\nu} T^a P_L b)
     \, G^a_{\mu\nu}\,.
\label{gmagnopsix} 
\end{array}
\end{equation}
The following four-quark (axial-)vector operators contribute:
\begin{equation}
\begin{array}{llll}
{\cal O}_{11,\tilde{g}}^q               \,= &\!  g_s^4(\mu) 
(\bar{s} \gamma_\mu  P_L b)\, 
(\bar{q} \gamma^\mu  P_L q) \,,           
               &  \quad \quad
{\cal O}_{11,\tilde{g}}^{q\,\prime}     \,= &\!  g_s^4(\mu)  
(\bar{s} \gamma_\mu  P_R b)\, 
(\bar{q} \gamma^\mu  P_R q) \,,                  \\[1.2ex] 
{\cal O}_{12,\tilde{g}}^q               \,= &\!  g_s^4(\mu) 
(\bar{s}_\alpha \gamma_\mu P_L b_\beta)\,
(\bar{q}_\beta \gamma^\mu P_L q_\alpha) \,,
               &  \quad \quad
{\cal O}_{12,\tilde{g}}^{q\,\prime}     \,= &\!  g_s^4(\mu) 
(\bar{s}_\alpha \gamma_\mu P_R b_\beta)\,
(\bar{q}_\beta \gamma^\mu P_R q_\alpha) \,,      \\[1.2ex]
{\cal O}_{13,\tilde{g}}^q               \,= &\!  g_s^4(\mu) 
(\bar{s} \gamma_\mu P_L b)\,
(\bar{q} \gamma^\mu P_R q) \,,
               &  \quad \quad
{\cal O}_{13,\tilde{g}}^{q\,\prime}     \,= &\!  g_s^4(\mu) 
(\bar{s} \gamma_\mu P_R b)\,
(\bar{q} \gamma^\mu P_L q) \,,                   \\[1.2ex]
{\cal O}_{14,\tilde{g}}^q               \,= &\!  g_s^4(\mu) 
(\bar{s}_\alpha \gamma_\mu P_L b_\beta)\,
(\bar{q}_\beta \gamma^\mu P_R q_\alpha) \,,
               &  \quad \quad
{\cal O}_{14,\tilde{g}}^{q\,\prime}     \,= &\!  g_s^4(\mu) 
(\bar{s}_\alpha \gamma_\mu P_R b_\beta)\,
(\bar{q}_\beta \gamma^\mu P_L q_\alpha) \,.     
\label{penguinboxop}
\end{array}
\end{equation}
where color indices are omitted for color-singlet currents.  These
operators arise 
from box diagrams through the exchange of two gluinos and from
penguin diagrams through the exchange of a gluino and a gluon.

In addition there are also the following 
four-quark (pseudo-)scalar operators which are induced at leading order by box diagrams with two 
gluino  exchanges 
 \begin{equation}
\begin{array}{llll}
{\cal O}_{15,\tilde{g}}^{q}         \,= &\!  g_s^4(\mu) 
(\bar{s} P_R b)\,
(\bar{q} P_R q)\,,
               &  \quad \quad
{\cal O}_{15,\tilde{g}}^{q\,\prime} \,= &\!  g_s^4(\mu) 
(\bar{s} P_L b)\,
(\bar{q} P_L q)\,,                         
               \\[1.2ex]
{\cal O}_{16,\tilde{g}}^{q}         \,= &\!  g_s^4(\mu) 
(\bar{s}_\alpha P_R b_\beta) \,
(\bar{q}_\beta  P_R q_\alpha)\,, 
               &  \quad \quad
{\cal O}_{16,\tilde{g}}^{q\,\prime} \,= &\!  g_s^4(\mu) 
(\bar{s}_\alpha P_L b_\beta) \,
(\bar{q}_\beta  P_L q_\alpha)\,,
               \\[1.2ex]
{\cal O}_{17,\tilde{g}}^{q}         \,= &\!  g_s^4(\mu) 
(\bar{s} P_R b)\,
(\bar{q} P_L q)\,,     
               &  \quad \quad
{\cal O}_{17,\tilde{g}}^{q\,\prime} \,= &\!  g_s^4(\mu) 
(\bar{s} P_L b)\,
(\bar{q} P_R q)\,,             
               \\[1.2ex]
{\cal O}_{18,\tilde{g}}^{q}         \,= &\!  g_s^4(\mu) 
(\bar{s}_\alpha P_R b_\beta) \,
(\bar{q}_\beta  P_L q_\alpha)\,,
               &  \quad \quad
{\cal O}_{18,\tilde{g}}^{q\,\prime} \,= &\!  g_s^4(\mu) 
(\bar{s}_\alpha P_L b_\beta) \,
(\bar{q}_\beta  P_R q_\alpha)\,,           
               \\[1.2ex]
{\cal O}_{19,\tilde{g}}^{q}         \,= &\!  g_s^4(\mu)  
(\bar{q}_\alpha P_R b_\beta)\,
(\bar{s}_\beta P_R q_\alpha)\,,
               &  \quad \quad
{\cal O}_{19,\tilde{g}}^{q\,\prime} \,= &\!  g_s^4(\mu) 
(\bar{q}_\alpha P_L b_\beta)\,
(\bar{s}_\beta  P_L q_\alpha)\,,
               \\[1.2ex]
{\cal O}_{20,\tilde{g}}^{q}         \,= &\!  g_s^4(\mu)  
(\bar{q} P_R b)\,
(\bar{s} P_R q)\,,
               &  \quad \quad
{\cal O}_{20,\tilde{g}}^{q\,\prime} \,= &\!  g_s^4(\mu) 
(\bar{q} P_L b)\,
(\bar{s} P_L q)\,.
\label{boxop}
\end{array} 
\end{equation}
These operators are shown to mix at  one-loop into the magnetic and 
chromomagnetic operators~\cite{Borzumati:1999qt} and, thus, contribute
to a LL analysis. Note that we defined ${\cal O}_{19,\tilde{g}}^{q}$ and 
${\cal O}_{20,\tilde{g}}^{q}$  (and also
${\cal O}_{19,\tilde{g}}^{q\,\prime}$,  
${\cal O}_{20,\tilde{g}}^{q\,\prime}$) 
 differently from
\cite{Borzumati:1999qt}. Denoting the operators from 
\cite{Borzumati:1999qt} by 
$\tilde{{\cal O}}_{19,\tilde{g}}^{q}$ 
and 
$\tilde{{\cal O}}_{20,\tilde{g}}^{q}$,  the transformation
between the operator bases reads:
\begin{equation}
{\cal O}_{19,\tilde{g}}^{q}=
-\frac{1}{2}\tilde{{\cal O}}_{15,\tilde{g}}^{q}
-\frac{1}{8}\tilde{{\cal O}}_{19,\tilde{g}}^{q}
\quad\text{and}\quad
{\cal O}_{20,\tilde{g}}^{q}=
-\frac{1}{2}\tilde{{\cal O}}_{16,\tilde{g}}^{q}
-\frac{1}{8}\tilde{{\cal O}}_{20,\tilde{g}}^{q}.
\end{equation}
Correspondingly we get for the Wilson coefficients:
\begin{align}
C_{15,\tilde{g}}^{q} &=
\tilde{C}_{15,\tilde{g}}^{q}
-4\tilde{C}_{19,\tilde{g}}^{q}\,, &
C_{16,\tilde{g}}^{q} &=
\tilde{C}_{16,\tilde{g}}^{q}
-4\tilde{C}_{20,\tilde{g}}^{q}\,, \nonumber\\
C_{19,\tilde{g}}^{q} &=
-8\tilde{C}_{19,\tilde{g}}^{q}\,, &
C_{20,\tilde{g}}^{q} &=
-8\tilde{C}_{20,\tilde{g}}^{q}.
\end{align}
Analogous relations are valid for the primed operators.

\section{Calculation of the Wilson coefficients of the magnetic\\ operators}\label{sec:magnetic}

\begin{figure}
\resizebox{\textwidth}{!}{\includegraphics{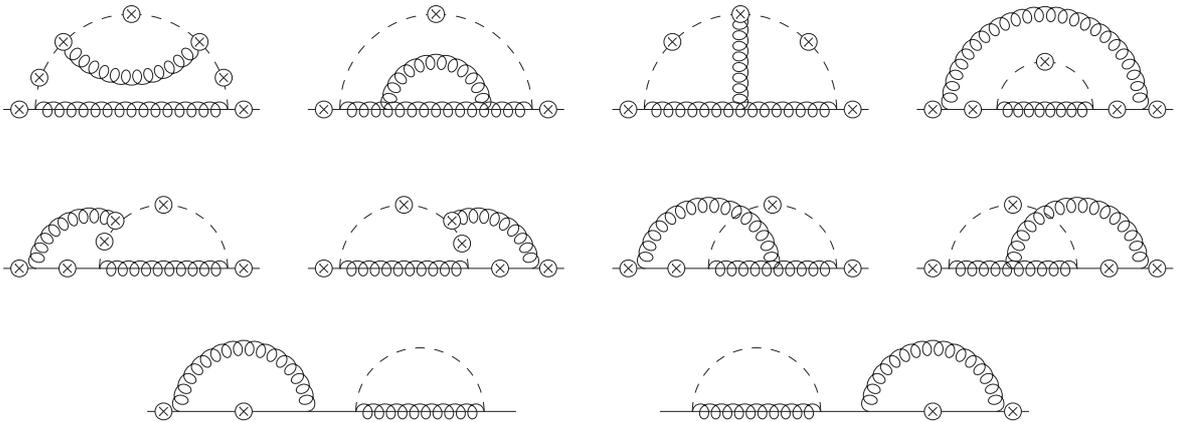}}
\caption{Gluon corrections: The solid, dashed, curly, and solid-curly lines represent quarks, squarks, gluons, and gluinos, respectively. The crosses denote the possible locations
  where the photon is emitted.}
\label{gluoncorrections}
\end{figure}
\begin{figure}
\resizebox{\textwidth}{!}{\includegraphics{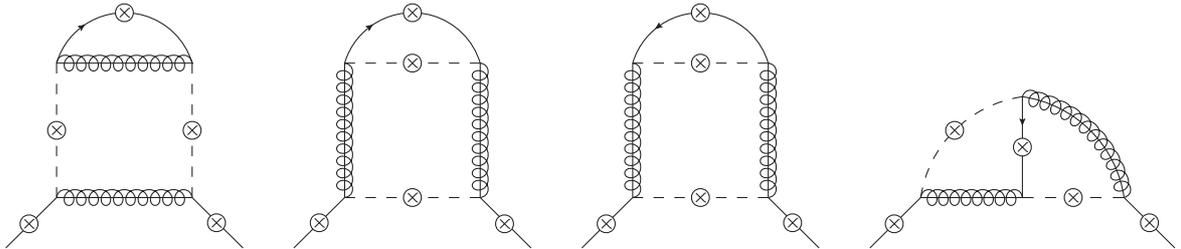}}
\caption{Gluino corrections: The crosses mark the places where a photon can be emitted.}
\label{gluinocorrections}
\end{figure}

\begin{figure}
\begin{center}
\resizebox{0.3\textwidth}{!}{\includegraphics{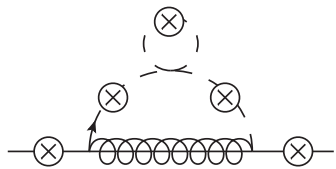}}
\caption{Diagrams containing a four-squark vertex.}
\label{four-squark_diag}
\end{center}
\end{figure}

The various Wilson coefficients $C_i$  appearing in the effective Hamiltonian 
$\mathcal{H}_{\text{eff}}^{\tilde{g}}$ can be expanded as follows:
\begin{equation}\begin{split}
C_i(\mu) = C_i^{(0)}(\mu) + \frac{\alpha_s(\mu)}{4\pi} \, C_i^{(1)}  + 
{\cal O}(\alpha_s^2) \, .
\end{split}\end{equation}

In order to extract the Wilson coefficients 
$C_{7\tilde{g},\tilde{g}}$
$C_{7b,\tilde{g}}$
$C_{7c,\tilde{g}}$
of
the magnetic operators 
${\cal O}_{7\tilde{g},\tilde{g}}$, 
${\cal O}_{7b,\tilde{g}}$  
${\cal O}_{7c,\tilde{g}}$ in   
\eqref{gmagnopfive} and \eqref{gmagnopsix},
we calculate the on-shell decay amplitude  $b\to s\gamma$
in the full theory  
and match the result with
the corresponding one obtained by  the effective Hamiltonian 
$\mathcal{H}_{\text{eff}}^{\tilde{g}}$ in
\eqref{HEffGluino}.
We require the $s$ quark, which we treat as a  massless particle, to be left-handed,
so  that only the unprimed operators are involved. The Wilson
coefficients of the primed operators (corresponding to a
right-handed $s$ quark) can be obtained at the very
end by interchanging $L$ and $R$ in
the results for the unprimed ones.

To get the LO pieces of the Wilson coefficients, we have to calculate
the amplitude ${\cal A}(b \to s \gamma)$ up to order $\alpha_s$.
The corresponding result in the full theory (expanded up to second
order in the external momenta and light masses) is given in Appendix
A. In the effective theory, only the matrix elements associated with 
the magnetic operators contribute at the lowest order of $\alpha_s$.
Consequently, we get for the leading-order Wilson coefficients
\begin{equation}\begin{split}
C_{7\tilde{g},\tilde{g}}^{(0)} = \left. K^{(0)}_{7\tilde{g},\tilde{g}}
\right|_{\epsilon\to0} \, , \qquad
C_{7b,\tilde{g}}^{(0)} = \left. K^{(0)}_{7b,\tilde{g}}
\right|_{\epsilon\to0} \, , \qquad 
C_{7c,\tilde{g}}^{(0)} =  0 \, ,
\end{split}\end{equation}
where the functions $K^{(0)}_{7\tilde{g},\tilde{g}}$ and
$K^{(0)}_{7b,\tilde{g}}$ are given in Appendix A.
They represent  the $d$-di\-men\-sional versions of the LO Wilson coefficients.

We now discuss the calculation of the NLO pieces 
$C_{7\tilde{g},\tilde{g}}^{(1)}$,
$C_{7b,\tilde{g}}^{(1)}$,
$C_{7c,\tilde{g}}^{(1)}$ of these Wilson coefficients.
There are two types of diagrams contributing to $b\to s\gamma$ in the full theory
at $\mathcal{O}(\alpha_s^2)$: 
\begin{description}
 \item[Full a)] Diagrams containing one virtual gluon and one
   gluino, shown in Figure~\ref{gluoncorrections}.
   We call these contributions ``gluon corrections''. 
 \item[Full b)] Contributions with two gluinos or diagrams containing
   squark tadpoles, depicted in Figures~\ref{gluinocorrections} and \ref{four-squark_diag}. We
   refer to these contributions as ``gluino corrections''.
\end{description}
As we will see, these two contributions can be renormalized separately in the full theory. 
It is therefore convenient to split also the
renormalized result obtained in the effective theory into two
contributions. This leads to a decomposition of the NLO pieces of the
Wilson coefficients of the magnetic operators into two pieces, as e.g.
$C_{7\tilde{g},\tilde{g}}^{(1)} =
C_{7\tilde{g},\tilde{g}}^{(1),a}+C_{7\tilde{g},\tilde{g}}^{(1),b}$.

The mentioned two contributions in the effective theory 
can be characterized as follows:
 
\begin{description}
 \item[Effective a)] Renormalized one-loop matrix elements $\langle s \gamma|C_i^{(0)}{\cal O}_i|b \rangle$ associated
with the ${\cal O}_7$- and ${\cal O}_8$-type operators;
 one-loop
matrix elements of the operators ${\cal O}^q_{11,\tilde{g}}$ to ${\cal
  O}^q_{14,\tilde{g}}$ (which are finite) weighted with the penguin part of the
corresponding LO Wilson coefficients (see  (31) in  \cite{Borzumati:1999qt});
tree-level
 matrix elements associated with the  operators 
\newline
$\frac{\alpha_s}{4\pi} \, C_{7i,\tilde{g}}^{(1),a} \, \langle s \gamma |{\cal O}_{7i,\tilde{g}}|b \rangle$, $i=b,\tilde{g}$.

 \item[Effective b)] 
 One-loop
matrix elements of the operators ${\cal O}^q_{11,\tilde{g}}$ to ${\cal
  O}^q_{14,\tilde{g}}$ (which are finite) weighted with the box part of the
corresponding LO Wilson coefficients (see  (32) in  \cite{Borzumati:1999qt});
renormalized one-loop
matrix elements of  $C^q_{15,\tilde{g}} \, {\cal
  O}^q_{15,\tilde{g}}$ to ${C^q_{20,\tilde{g}} \, \cal
  O}^q_{20,\tilde{g}}$  (see  (33) in  
\cite{Borzumati:1999qt});
tree-level
matrix elements associated with the  operators 
$\frac{\alpha_s}{4\pi} \, C_{7i,\tilde{g}}^{(1),b} \, \langle s \gamma |{\cal O}_{7i,\tilde{g}}|b \rangle,\,\,\,
i=b,c,\tilde{g}$.
\end{description}
For the precise definition of these two parts, see (\ref{Arenaeff}) and (\ref{Abrenaeff}).

In the following, we separately work out  $C^{(1),a}_7$ and $C^{(1),b}_7$
using dimensional regularization (DREG) both on the full theory and the
effective theory side. Later we will take into account the important point, that one should use
dimensional reduction (DRED) in the full theory to preserve
supersymmetry. This is done  in Section \ref{sec:DRbarTrans} by appropriately shifting the strong coupling  constant $g_s$ and the gluino mass in the leading order contributions of
the full theory.

\subsection{Calculation of $C^{(1),a}_{7\tilde{g},\tilde{g}}$ and $C^{(1),a}_{7b,\tilde{g}}$}

\subsubsection{Full theory}

To get the order $\alpha_s^2$ corrections to the decay amplitude
corresponding to the contributions of class~a) in the full theory, 
we expand the diagrams shown in Figure~\ref{gluoncorrections}  in inverse
powers of the heavy mass $M$ ($M=m_{\tilde{g}}$ or $M=m_{\tilde{q}}$),
using the hard-mass procedure~\cite{Smirnov:2002pj}.
We systematically retain all terms up to order $1/M^2$. When
evaluating a genuine two-loop diagram in
Figure~\ref{gluoncorrections} according to this method, we 
get two contributions: The so-called  hard contribution which is
obtained by a naive Taylor 
  expansion of the two-loop integral with respect to the external
momenta, and the 
so-called soft contribution (for details how to get this contribution,
see \cite{Smirnov:2002pj}). 
Adding the hard contributions of all
diagrams defines the hard part $iA^{bare}_{a,full,h}$, while the sum of the
soft contributions defines the part
$iA_{a,full,s}$, so that
\begin{equation}\begin{split}
iA^{bare}_{a,full}=iA^{bare}_{a,full,h} +i A^{bare}_{a,full,s} \, .
\end{split}\end{equation}
Note that in this section the symbol $A$ is used for the contribution of
order $\alpha_s^2$ of the relevant amplitude.
Instead of calculating the soft part $iA^{bare}_{a,full,s}$ 
according to the rules of hard-mass procedure, it can be generated using
the effective theory at LO (as we checked explicitly) in the following way:
\begin{equation}
\label{abareafulls}
\begin{split}
i \, A^{bare}_{a,full,s} = & K^{(0)}_{7b,\tilde{g}} \, \langle s \gamma |{\cal O}_{7b,\tilde{g}}|b\rangle_{\mbox{\tiny 1-loop}}
              +K^{(0)}_{7\tilde{g},\tilde{g}} \, \langle s \gamma |{\cal
                O}_{7\tilde{g},\tilde{g}} |b\rangle_{\mbox{\tiny 1-loop}}  \\
              & +K^{(0)}_{8b,\tilde{g}} \, \langle s \gamma |{\cal O}_{8b,\tilde{g}}|b\rangle_{\mbox{\tiny 1-loop}}+
              K^{(0)}_{8\tilde{g},\tilde{g}} \, \langle s \gamma |{\cal
                O}_{8\tilde{g},\tilde{g}}|b\rangle_{\mbox{\tiny
                  1-loop}}  \\
              & +\sum_{i=11}^{14} C^{q,a}_{i,\tilde{g}} \, \langle s \gamma |{\cal O}^q_{i,\tilde{g}}|b\rangle_{\mbox{\tiny 1-loop}}
              \\
              & +(Z_{m_b}^{OS,a}-1) \, K^{(0)}_{7b,\tilde{g}} \, \langle s \gamma |{\cal
                O}_{7b,\tilde{g}}|b\rangle_{\mbox{\tiny tree}} \, .
\end{split}
\end{equation}
The various ($d$-dimensional Wilson coefficient) functions $K^{(0)}$, 
which in the limit $\epsilon \to 0$ coincide with the correponding
ordinary LO Wilson coefficents $C^{(0)} $, are given in Appendix~A.
$C^{q,a}_{i,\tilde{g}}$ denote the penguin parts of the Wilson
coefficients $C^{q}_{i,\tilde{g}}$ ($i=11-14$); they are explicitly given in (31)
of \cite{Borzumati:1999qt}.
Note that
it is important to keep the functions $K^{(0)}$ in $d=4-2\epsilon$
dimensions, because they are multiplied with one-loop matrix elements
which are singular in $\epsilon$. On the other hand, the matrix
elements of the operators ${\cal O}^q_{i,\tilde{g}}$ $(i=11-14)$ are
finite, therefore it is sufficient to weight them with the corresponding (4-dimensional) 
Wilson coefficients  
$C^{q,a}_{i,\tilde{g}}$. Finally, 
$Z_{m_b}^{OS,a}$ denotes the renormalization constant of $m_b$ due
to gluonic corrections in the on-shell scheme:
\begin{equation}\begin{split}
Z_{m_b}^{OS,a} = 1 - \frac{\alpha_s}{4\pi} \, C_F \,
\left[\frac{3}{\epsilon} +3 \ln \frac{\mu^2}{m_b^2} +4  \right] \, ,
\end{split}\end{equation}
with  $C_F=4/3$.

The renormalized amplitude $iA^{ren}_{a,full}$ in the full theory is obtained by adding
the counter\-term $iA^{ct}_{a,full}$ induced by renormalizing the
following quantities in the leading order expression for the amplitude
given in  Appendix A:
The gluino-mass $m_{\tilde{g}}$,  the down-type squark masses $m_{\tilde{d}_k}$  and the strong
coupling constant of Yukawa type $g_{s,Y}$. Furthermore,   the on-shell renormalization
constants $\sqrt{Z_{2b}^{OS,a}}$ and $\sqrt{Z_{2s}^{OS,a}}$ have to be
attached. 
As we are discussing gluon-corrections in this section, we only take
into account the gluonic contributions of the involved renormalization
constants. The counterterm amplitude then reads 
\begin{equation}
\begin{split}
\label{ActAfull}
i \, A^{ct}_{a,full} = & \left( 2 \, \delta Z^a_{g_{s,Y}} + \frac{1}{2}
  \, \delta
  Z_{2b}^{OS,a} +\frac{1}{2} \, \delta Z_{2s}^{OS,a} \right) \, \, \left(
K^{(0)}_{7b,\tilde{g}} \,
\langle s \gamma |{\cal O}_{7b,\tilde{g}}|b\rangle_{\mbox{\tiny tree}}  \right.  \\
& +\left. K^{(0)}_{7\tilde{g},\tilde{g}} \,
\langle s \gamma |{\cal O}_{7\tilde{g},\tilde{g}}|b\rangle_{\mbox{\tiny tree}} \right) +
R^a_{\tilde{g}\tilde{d}_k} \, .
\end{split}
\end{equation}
$Z^a_{g_{s,Y}}$, derived in the  Appendix~\ref{gYren}, reads
\begin{equation}
Z^a_{g_{s,Y}} = 1 - \frac{\alpha_s}{4\pi} \,
\left[\frac{3(C_A+C_F)}{2\epsilon}\right] \, ,
\end{equation}
with $C_A=3$, while the on-shell renormalization constants for the $b$
quark and the
massless $s$ quark, which will drop out in the matching equation
(\ref{matchingequation}),  read (see
Appendix~\ref{sec:quarkrenormalization}, \eqref{quarkZ}) 
\begin{equation}
\begin{split}
 Z_{2b}^{OS,a} &=\, 1 - \, \frac{\alpha_s}{4\pi} \,C_F \, \left[\frac{3}{\epsilon}
   + 4 + 3 \, \ln \frac{\mu^2}{m_b^2}
   \right] , \\
 Z_{2s}^{OS,a} &=\,0. \\
\end{split}
\end{equation}  
In (\ref{ActAfull}), the contribution $R^a_{\tilde{g}\tilde{d}_k}$ is
obtained by shifting (according to the $\overline{\mbox{MS}}$ scheme)
\begin{equation}
\begin{split}
 m_{\tilde{g}}& \to m_{\tilde{g}} \, \left[ 1 - \frac{\alpha_s}{4\pi} \,
  C_A \, \frac{3}{\epsilon} \right]\,,  \\
 m^2_{\tilde{d}_k}& \to m^2_{\tilde{d}_k} \, \left[ 1 -
  \frac{\alpha_s}{4\pi} \, C_F \, \frac{3}{\epsilon} \right]
 \,,
\end{split}
\end{equation}
in the leading order expression for $i \, A_{full}$ in Appendix A,
followed by an expansion in $\alpha_s$ and keeping the term proportional
to $\alpha_s^2$.

Finally, putting everything together, the renormalized result in the full theory reads 
\begin{equation}
\begin{split}
\label{Arenafull}\\
i \, A^{ren}_{a,full} = & i \, A^{bare}_{a,full,h} +  i \, A^{bare}_{a,full,s} \\ 
&+ \left[ 2 \, \delta Z^a_{g_{s,Y}} + \frac{1}{2}
  \, \delta
  Z_{2b}^{OS,a} +\frac{1}{2} \, \delta Z_{2s}^{OS,a} \right] \\
 &  \times \, \left(
K^{(0)}_{7b,\tilde{g}} \,
\langle s \gamma |{\cal O}_{7b,\tilde{g}}|b\rangle_{\mbox{\tiny tree}}  +  K^{(0)}_{7\tilde{g},\tilde{g}} \,
\langle s \gamma |{\cal O}_{7\tilde{g},\tilde{g}}|b\rangle_{\mbox{\tiny tree}} \right) +
R^a_{\tilde{g}\tilde{d}_k} \, .
\end{split}
\end{equation} 
We now turn to the effective theory side. 

\subsubsection{Effective theory}
After converting the running
$b$-quark mass present in the operator ${\cal O}_{7b,\tilde{g}}$ to
the pole mass, the renormalized amplitude
$iA^{ren}_{a,eff}$ can be written as
\begin{equation}
\begin{split}
\label{Arenaeff}
i \, A^{ren}_{a,eff} = & \frac{\alpha_s}{4\pi} \, C_{7g,g}^{(1),a} \,
\langle s \gamma |{\cal O}_{7\tilde{g},\tilde{g}}|b\rangle_{\mbox{\tiny tree}}
+ \frac{\alpha_s}{4\pi} \, C_{7b,g}^{(1),a} \,
\langle s \gamma |{\cal O}_{7b,\tilde{g}}|b\rangle_{\mbox{\tiny tree}}  
\\
&+ K^{(0)}_{7b,\tilde{g}} \, \langle s \gamma |{\cal O}_{7b,\tilde{g}}|b\rangle_{\mbox{\tiny 1-loop}}+
              K^{(0)}_{7\tilde{g},\tilde{g}} \, \langle s \gamma |{\cal
                O}_{7\tilde{g},\tilde{g}} |b\rangle_{\mbox{\tiny 1-loop}} \\
              &+ K^{(0)}_{8b,\tilde{g}} \, \langle s \gamma |{\cal O}_{8b,\tilde{g}}|b\rangle_{\mbox{\tiny 1-loop}}+
              K^{(0)}_{8\tilde{g},\tilde{g}} \, \langle s \gamma |{\cal
                O}_{8\tilde{g},\tilde{g}}|b\rangle_{\mbox{\tiny 1-loop}} \\
              &+ \sum_{i=11}^{14} C^{q,a}_{i,\tilde{g}} \, \langle s \gamma |{\cal O}^q_{i,\tilde{g}}|b\rangle_{\mbox{\tiny 1-loop}}
              \\
              &+ (Z_{m_b}^{OS,a}-1) \, K^{(0)}_{7b,\tilde{g}} \, \langle s \gamma |{\cal
                O}_{7b,\tilde{g}}|b\rangle_{\mbox{\tiny tree}} \,   \\
              &+ \left[ 2 \, \delta Z_{g_{s}} + \delta
               \,  Z_{77,b\tilde{g}} + \frac{1}{2} \, \delta
               Z_{2b}^{OS,a} +\frac{1}{2} \, \delta Z_{2s}^{OS,a} \right] \, 
     K^{(0)}_{7b,\tilde{g}} \, \langle s \gamma |{\cal
       O}_{7b,\tilde{g}}|b\rangle_{\mbox{\tiny tree}}   \\
              &+ \left[ 2 \, \delta Z_{g_{s}} + \delta
               \,  Z_{77,\tilde{g}\tilde{g}} + \frac{1}{2} \, \delta
               Z_{2b}^{OS,a} +\frac{1}{2} \, \delta Z_{2s}^{OS,a} \right] \, 
     K^{(0)}_{7\tilde{g},\tilde{g}} \, \langle s \gamma |{\cal
       O}_{7\tilde{g},\tilde{g}}|b\rangle_{\mbox{\tiny tree}}  \\
         & + \delta \,  Z_{87,\tilde{g}\tilde{g}}  \, 
     K^{(0)}_{8g,\tilde{g}} \, \langle s \gamma |{\cal
       O}_{7\tilde{g},\tilde{g}}|b\rangle_{\mbox{\tiny tree}} + 
           \delta \,  Z_{87,b\tilde{g}}  \, 
     K^{(0)}_{8b,\tilde{g}} \, \langle s \gamma |{\cal
       O}_{7b,\tilde{g}}|b\rangle_{\mbox{\tiny tree}}  \, ,
\end{split}
\end{equation}
where the terms in the lines 2--5 correspond to $i \,
A^{bare}_{a,full,s}$ in (\ref{abareafulls}) and will cancel in  the
matching equation (\ref{matchingequation}).
Those renormalization constants, whose effects do not drop out
in the matching step (\ref{matchingequation}), read
\begin{align}
\begin{split}
 Z_{g_s} = 1- \frac{\alpha_s}{4\pi} \, \frac{23}{6 \, \epsilon} \, ,& \qquad
Z_{77,\tilde{g}\tilde{g}} = 1 + \frac{\alpha_s}{4\pi} \, \frac{9}{\epsilon} \,
, \qquad
 \\
Z_{77,b\tilde{g}} = 1 + \frac{\alpha_s}{4\pi} \, \frac{13}{\epsilon} \, ,& \qquad
Z_{87,\tilde{g}\tilde{g}} = Z_{87,b\tilde{g}} = 
1 - \frac{\alpha_s}{4\pi} \, \frac{16}{9 \, \epsilon} \, .
\end{split}
\end{align}
Note that in (\ref{Arenaeff}) we put the $d$-dimensional versions
$K^0_{7\tilde{g},\tilde{g}}$, $K^0_{7b,\tilde{g}}$,
$K^0_{8\tilde{g},\tilde{g}}$ and $K^0_{8b,\tilde{g}}$   
of the LO Wilson coefficients, which implies that
the  infrared structure on the full and effective theory side are identical.
Moreover, the use of the $d$-dimensional versions of the Wilson
  coefficients  in the
 effective theory  also implies that all contributions due to the
 on-shell renormalization 
constants of the $b$ and $s$ quark will cancel out in the matching
condition 
(\ref{matchingequation}). These features {\it manifestly} reflect  the
fact  that the 
determination of the {\it short-distance}  Wilson coefficients is  
independent of the infrared structure and 
of the choice of the external states.

\subsubsection{Extracting $C^{(1),a}_{7\tilde{g},\tilde{g}}$ and
  $C^{(1),a}_{7b,\tilde{g}}$}\label{sec:Extracta}
The coefficients $C^{(1),a}_{7\tilde{g},\tilde{g}}$ and
  $C^{(1),a}_{7b,\tilde{g}}$ are obtained by requiring that
\begin{equation}
\label{matchingequation}
 iA^{ren}_{a,full} = iA^{ren}_{a,eff} \, .
\end{equation}
Using
$x_{d_k} =m_{\tilde{d}_k}^2/m_{\tilde{g}}^2$ and
 $L_\mu=\ln(\mu^2/m^2_{\tilde{g}})$, we get
\begin{equation}
\begin{split}
\label{resultpart1}
& C^{(1),a}_{7\tilde{g},\tilde{g}} =
\frac{4}{3} \, \frac{1}{16\pi^2} \, \frac{1}{m_{\tilde{g}}} \, 
\sum\limits_{k=1}^6 \Gamma_{DR}^{kb} \Gamma_{DL}^{ks*} \, 
 h^{(1),a}_{7\tilde{g},\tilde{g}}(x_{d_k}),  \\
& C^{(1),a}_{7b,\tilde{g}} =
- \frac{4}{3} \, \frac{1}{16\pi^2} \, \frac{1}{m_{\tilde{g}}^2} \, 
\sum\limits_{k=1}^6 \Gamma_{DL}^{kb} \Gamma_{DL}^{ks*} \, 
h^{(1),a}_{7b,\tilde{g}}(x_{d_k}),  
\end{split}
\end{equation}
with 
\begin{equation}\begin{split}
 h^{(1),a}_{7\tilde{g},\tilde{g}}(x) =& \left( -\frac{55 x^2+69 x+2}{9 (x-1)^3} + \frac{2 x \left(9 x^2+40 x+14\right)}{9 (x-1)^4} \ln x \right)L_\mu + \frac{x \left(x^2-11 x+13\right)}{18 (x-1)^4} \ln^2 x \\
 &+ \frac{19 x^2+60 x-15}{9 (x-1)^3} \mbox{Li}_2(1-x) + \frac{28 x^3+120 x^2-15 x-1}{9 (x-1)^4} \ln x\\
 &+ \frac{-14 x^2-333 x+83}{18 (x-1)^3},\\[2mm]
 h^{(1),a}_{7b,\tilde{g}}(x) =&\left(\frac{-5 x^3+384 x^2+609 x+20}{108 (x-1)^4} -\frac{x \left(18 x^2+107 x+43\right)}{18 (x-1)^5} \ln x\right)L_\mu + \frac{x (x+11)}{36 (x-1)^4} \ln^2 x\\
 &+ \frac{-17 x^2-86 x+15}{18 (x-1)^4} \mbox{Li}_2(1-x) + \frac{87 x^4-537 x^3-2997 x^2+265 x+14}{324 (x-1)^5}\ln x \\
 &+ \frac{-799 x^3+1719 x^2+10431 x-1847}{972 (x-1)^4}.
\end{split}\end{equation}
           
Our results are in full agreement with those of
Bobeth et al. \cite{Bobeth:1999ww} who matched the corresponding
off-shell Greens functions. 

Finally, we emphasize that one could also insist to use the four-dimensional version
of the Wilson coefficients in the effective theory, because one can  work out the explicit infrared~(IR) structure 
of the amplitudes in the effective and full theory by distinguishing  IR poles ($1/\epsilon_{\rm{ir}}$) from ultraviolet (UV) poles  ($1/\epsilon$).
One then finds 
that the IR structures in $iA^{ren}_{a,full}$ and
$iA^{ren}_{a,eff}$ are of the form 
\[
\left( \frac{A}{\epsilon^2_{\rm{ir}}} + \frac{B}{\epsilon_{\rm{ir}}}
\, \right) \, K^{(0)}_{7} \, \langle O_7 \rangle_{\mbox{\tiny tree}} 
\quad \mbox{and} \quad 
\left( \frac{A}{\epsilon^2_{\rm{ir}}} + \frac{B}{\epsilon_{\rm{ir}}}
\, \right) \, C^{(0)}_{7}  \,  \langle O_7 \rangle_{\mbox{\tiny tree}} \, ,
\]
respectively. In both cases the same singular factor multiplies the
lowest order amplitude of the full theory and the effective theory,
respectively. On general grounds, the corresponding  statement about the IR structure
holds  for the bremsstrahlung contributions. So the correct matching conditions,  
to extract NLO pieces of the Wilson coefficients
$C^{(1),a}_{7\tilde{g},\tilde{g}}$ and $C^{(1),a}_{7b,\tilde{g}}$,
can be achieved  in this case by discarding these explicitly identified  
IR sensitive terms in $iA^{ren}_{a,full}$ and
$iA^{ren}_{a,eff}$.

\subsection{Calculation of $C^{(1),b}_{7\tilde{g},\tilde{g}}$, $C^{(1),b}_{7b,\tilde{g}}$, and $C^{(1),b}_{7c,\tilde{g}}$}
\subsubsection{Full theory}
\begin{figure}
\begin{center}
\resizebox{0.5\textwidth}{!}{\includegraphics{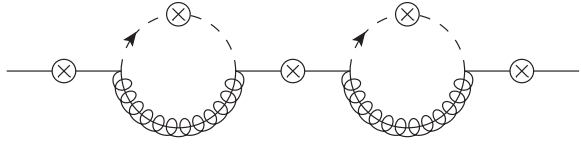}}
\caption{One-particle reducible diagrams, which are taken into account
through the quark-field renormalization constants}\label{onepartred}
\end{center}
\end{figure}

In order to get the full theory NLO contributions to the decay
amplitude belonging to class
b), we have to calculate 
irreducible two-loop diagrams with two virtual gluinos depicted in
Figure~\ref{gluinocorrections} 
and contributions with squark tadpoles shown in
Figure~\ref{four-squark_diag}.
We denote the contributions of these diagrams
by $i A_{b,irred,full}^{bare}$.
Additionally, there are reducible diagrams shown in
Figure~\ref{onepartred} leading to a contribution to the decay
amplitude denoted $i A_{b,red,full}^{bare}$.

First, we discuss the calculation of the irreducible diagrams.
We expand the corresponding two-loop diagrams according to the rules
of the hard-mass procedure \cite{Smirnov:2002pj} in 
inverse powers of the heavy masses, which are in our application $m_{\tilde{q}_k}$,
$m_{\tilde{g}}$ and $m_t$. The class of diagrams containing
squark tadpoles 
(Figure~\ref{four-squark_diag}) does  not contain any light propagators
and can
therefore just be naively Taylor-expanded in the external momenta. The
diagrams 
with two virtual gluinos (Figure~\ref{gluinocorrections}) can also
contain 
propagators of light quarks. Thus, besides the naively expanded two-loop diagrams, we 
also have to take into account  contributions with one-loop
sub-diagrams containing all heavy lines which have to be expanded in
the external momenta. The result is inserted in the second loop which
contains only light propagators as an effective vertex.
Logarithms of the small masses are
only generated in the latter contributions and cancel in the matching
procedure against equal terms on
the effective side stemming from matrix elements of the four-Fermi
operators ${\cal O}^q_{11,\tilde{g}}$ to ${\cal O}^q_{20,\tilde{g}}$. 

The integrals which result  from the naive Taylor expansion  are
two-loop vacuum integrals.
Using integration-by-parts identities~\cite{Chetyrkin:1981qh,Tkachov:1981wb}
and partial fraction decompositions, we can reduce all these integrals
to the following master integrals:
\begin{equation}
\begin{split}
&\int\frac{d^dk}{(2\pi)^d}\frac{1}{k^2-m^2} =  
\frac{-i}{(4\pi)^{\frac{d}{2}}}\Gamma\left(1-\frac{d}{2}\right)m^{d-2},\\[2mm]
&\int\frac{d^dk_1}{(2\pi)^d}\frac{d^dk_2}{(2\pi)^d}
\frac{1}{(k_1^2-m_1^2) \, (k_2^2-m_2^2) \, [(k_1+k_2)^2-m_3^2]} = 
\frac{1}{512\pi^4}m_3^{2-4\epsilon}
\frac{\Gamma(1+\epsilon)^2}{(1-\epsilon)(1-2\epsilon)}\\
&
\times\bigg[
-\frac{1+x+y}{\epsilon^2}+\frac{2(x\ln x+y\ln y)}{\epsilon}
-x\ln^2x-y\ln^2y+(1-x-y)\ln x\ln y-
\lambda^2\Phi(x,y)
\bigg],
\end{split}
\end{equation}
where $d=4-2\epsilon$, $\lambda^2=(1-x-y)^2-4xy$, $x=m_1^2/m_3^2$, and
$y=m_2^2/m_3^2$. The function $\Phi(x,y)$ has been evaluated in
\cite{Davydychev:1992mt}. 
For $\lambda^2>0$ it is represented by dilogarithms as
\begin{equation}
\begin{split}
\Phi(x,y) = \frac{1}{\lambda}
&\bigg[2\ln\frac{1+x-y-\lambda}{2}
\ln\frac{1-x+y-\lambda}{2}-\ln x\ln y
\\
&-2\text{Li}_2\frac{1+x-y-\lambda}{2}
-2\text{Li}_2\frac{1-x+y-\lambda}{2}+
\frac{\pi^2}{3}
\bigg],
\end{split}
\end{equation}
whereas for $\lambda^2<0$ it can be written in terms of Clausen's
functions:
\begin{equation}
\begin{split}
\Phi(x,y) =& \frac{2}{\sqrt{-\lambda^2}}
\bigg[
\text{Cl}_2\left(2\arccos\frac{-1+x+y}{2\sqrt{xy}}\right)+
\text{Cl}_2\left(2\arccos\frac{1+x-y}{2\sqrt{x}}\right)
\\
&+\text{Cl}_2\left(2\arccos\frac{1-x+y}{2\sqrt{y}}\right)
\bigg].
\end{split}
\end{equation}

We now turn to the reducible diagrams. Their contribution to the decay 
amplitude can be obtained by amputating the external fermion legs
(with squark-gluino self-energy insertions) followed by attaching the
LSZ-factor, i.e. the gluino parts of the renormalization constants of the $b$-
and $s$-quark fields in the on-shell scheme which are derived in detail
in Appendix~\ref{sec:quarkrenormalization}. 
This amounts to replace the $\Gamma_{DL/R}$ matrices in the
LO amplitude, given in Appendix~\ref{sec:LOresult}, according to

\begin{equation}\label{substitution}
 \Gamma_{DL/R}^{ki}\rightarrow \Gamma_{DL/R}^{kj} \, \frac{1}{2} \, \delta
 Z^{b,L/R\phantom{*}}_{ji} = 
\Gamma_{DL/R}^{kj} \, \frac{1}{2} \,  \left(\delta Z^{b,L/R,H}_{ji} +
\delta Z^{b,L/R,AH}_{ji}\right), 
\end{equation}
where a summation over the index $j$ is understood. 

In (\ref{substitution}), we split the renormalization constants into a hermitian and an antihermitian part as
\begin{equation}
 \delta Z^{b,L/R} = \delta Z^{b,L/R,H} + \delta Z^{b,L/R,AH}.
\end{equation}
Explicitly, they read in the off-diagonal case (using $x_{d_k}=m_{\tilde{d}_k}^2/m_{\tilde{g}}^2$)
\begin{equation}
 \begin{split}
 \delta Z_{ji}^{b,L,AH}=& \frac{\alpha_s}{4 \pi}
 \frac{C_F}{m_i^2-m_j^2} \sum\limits_{k=1}^6 
\Biggl\{(\Gamma_{DL}^{ki} \Gamma_{DR}^{kj*} m_j+\Gamma_{DL}^{kj*}
\Gamma_{DR}^{ki} m_i) 
\left(\frac{m_i^2+m_j^2}{m_{\tilde{g}} }f_1(x_{d_k})-4 m_{\tilde{g}}f_2(x_{d_k})\right) \Biggr.\\
  &\Biggl. +\left(\Gamma_{DL}^{ki} \Gamma_{DL}^{kj*} \left(m_i^2+m_j^2\right)+2 \Gamma_{DR}^{ki} \Gamma_{DR}^{kj*} m_i
   m_j\right)f_3(x_{d_k}) \Biggr\} \, ,
 \end{split}
\end{equation}
\begin{equation}
 \begin{split}
 \delta Z_{ji}^{b,L,H}=&\frac{\alpha_s}{4 \pi} C_F \sum\limits_{k=1}^6
 \left(\frac{1}{m_{\tilde{g}}} 
(\Gamma_{DL}^{kj*} \Gamma_{DR}^{ki} m_i + \Gamma_{DL}^{ki}
 \Gamma_{DR}^{kj*} m_j) 
f_1(x_{d_k}) + \Gamma_{DL}^{ki} \Gamma_{DL}^{kj*} f_3(x_{d_k})
 \right) \, ,
 \end{split}
\end{equation}
where $m_i$ are the masses of the down-type quarks. The
corresponding results for the
 right-handed versions are obtained by interchanging the labels 
$R$ and $L$. The functions $f_i(x)$ are given in
Appendix~\ref{sec:quarkrenormalization}. In the diagonal case, we get
(using $L_\mu = \ln(\mu^2/m^2_{\tilde{g}})$)
\begin{equation}\begin{split}
\delta Z_{ii}^{b,L,AH}=&\,0, \\
\delta Z_{ii}^{b,L,H}=&\frac{\alpha_s}{4 \pi} C_F \sum\limits_{k=1}^6
    \left(\frac{m_i}{m_{\tilde{g}}}
 (\Gamma_{DL}^{ki*} \Gamma_{DR}^{ki} + \Gamma_{DL}^{ki}
    \Gamma_{DR}^{ki*} )
 f_1(x_{d_k}) + \Gamma_{DL}^{ki} \Gamma_{DL}^{ki*} f_3(x_{d_k})
 \right) \\
&- \frac{\alpha_s}{4 \pi} C_F \left( \frac{1}{\epsilon} +L_\mu \right)
\, .  
\end{split}\end{equation}

Performing the shifts (\ref{substitution}) in the LO result yields
$i A^{bare}_{b,red,full}$ at the NLO level. 
Since the hermitian parts of the $Z$ factors do not contain any
inverse powers of the light quark masses, the corresponding
contribution to $i A^{bare}_{b,red,full}$ can be expanded up to linear
order in $m_b$. The contributions proportional to $m_b^0$ and $m_b^1$
will end up in the Wilson coefficients of the 
magnetic operators of dimension five and six, respectively.

For the contributions induced by the antihermitian parts of the
$Z$-factors this splitting  does not work
since they also contain inverse powers of the light quark masses,
i.e. {\it chirally enhanced} terms which have been recently discussed in \cite{Crivellin:2009ar}.   Instead, we proceed in such a way that 
the contributions involving the LO functions $K^{(0)}_{7\tilde{g},\tilde{g}}$ and $K^{(0)}_{7b,\tilde{g}}$
will end up in the Wilson coefficients of the 
magnetic operators of dimension five and six, respectively.
This ends the discussion for constructing
$iA^{bare}_{b,red,full}$.
%
Note that in principle one could renormalize
the $\Gamma$ matrices in a way that the antihermitian parts of the
quark field renormalization constants get exactly
cancelled. 
However, 
the terms induced by the antihermitian parts of quark field
renormalization constants remain in the decay amplitude in our scheme
in which we renormalize the
$\Gamma$ matrices minimally (see counterterms induced by the
renormalization of the $\Gamma$ matrices below and also
Appendix~\ref{gYren}).

We now discuss the counterterm contribution $iA^{ct}_{b,full}$ which
gets induced by renormalizing
the strong coupling constant of Yukawa-type $g_{s,Y}$, 
the gluino mass $m_{\tilde{g}}$, the squark masses $m_{\tilde{d}_k}$, 
and the $\Gamma_{D,L/R}$ matrices
in the LO amplitude. We get
\begin{equation}
\begin{split}
\label{ActBfull}
i \, A^{ct}_{b,full} = &  2 \,  \delta Z^b_{g_{s,Y}} \, \, \left(K^{(0)}_{7b,\tilde{g}} \,
\langle s \gamma |{\cal O}_{7b,\tilde{g}}|b\rangle_{\mbox{\tiny tree}} +  K^{(0)}_{7\tilde{g},\tilde{g}} \,
\langle s \gamma |{\cal
  O}_{7\tilde{g},\tilde{g}}|b\rangle_{\mbox{\tiny tree}} \right) + 
R^b_{\tilde{g}\tilde{d}_k\Gamma} \, .
\end{split}
\end{equation}
$\delta Z^b_{g_{s,Y}}$ is derived in Appendix~\ref{gYren} and reads
\begin{equation}
 \delta Z_{g_s,Y}^{b} = \frac{\alpha_s}{4\pi} \frac{1}{\epsilon}\left( \frac{3}{2} C_F+ tr\, n_f \right),
\end{equation}
with $tr=1/2$ and $n_f=6$ denoting the number of flavors.
The term $R^b_{\tilde{g}\tilde{d}_k\Gamma}$ is obtained by replacing 
\begin{equation}
  m_{\tilde{g}} \to m_{\tilde{g}} \left[1 +\frac{\alpha_s}{4\pi} \, 2 \, tr \,
\frac{n_f}{\epsilon}  \right],
\end{equation}
\begin{equation}\label{squarkmassrenb}
 \begin{split}
  m_{\tilde{d}_k}^2 \to m_{\tilde{d}_k}^2 &+ \frac{\alpha_s}{\pi} C_F \, \left[
  \sum_d \left(
m_{\tilde{g}} \, m_d \, (\Gamma_{DR}^{kd*} \Gamma_{DL}^{kd}+ \Gamma_{DL}^{kd*}
\Gamma_{DR}^{kd}) \right) \right] \frac{1}{\epsilon}    
+ \frac{\alpha_s}{2\pi} C_F \, m_{\tilde{g}}^2 \,  (x_{d_k}-2) \frac{1}{\epsilon}\\
&
+ \frac{\alpha_s}{3\pi} \, m_{\tilde{g}}^2 \,X_D^{kl} \, x_{d_l} \, X_D^{lk} \,  \frac{1}{\epsilon},
 \end{split}
\end{equation}
and
\begin{equation}
 \Gamma_{DL/R}^{ki} \to \Gamma_{DL/R}^{ki} + \delta\Gamma_{D}^{kl} \, 
\Gamma_{DL/R}^{li},
\end{equation}
in the leading order expression for $i A_{full}$ in
Appendix~\ref{sec:LOresult}, followed by 
an expansion in $\alpha_s$ and keeping the term proportional to $\alpha_s^2$.
The explicit sum in \eqref{squarkmassrenb} runs over all down-type quark
flavors of mass $m_d$, while
the one over $l=1,...,6$ is implicitly understood. 
Again, $x_{d_k} = m^2_{\tilde{d}_k}/m^2_{\tilde{g}}$ and the 
matrix $X_D$ is given by
\begin{equation}
 X_D^{kk'} =
\Gamma_{DR}^{ki} \Gamma_{DR}^{k'i*} - 
\Gamma_{DL}^{ki} \Gamma_{DL}^{k'i*}, 
\end{equation}
with $i$ summed over $i=1,2,3$. The shift in the coupling matrices,
described by the matrix 
$\delta\Gamma_{D}^{kl}$ is derived 
in Appendix~\ref{gYren}  and reads
\begin{equation}
\begin{split}
 \delta\Gamma_{D}^{kl}= - \frac{\alpha_s}{\pi} \,
\frac{1}{x_{d_l}-x_{d_k}} \, &\left\{
 C_F \, \left[ \sum_d \left(
\frac{m_d}{m_{\tilde{g}}} \, (\Gamma_{DL}^{kd} \Gamma_{DR}^{ld*} + 
\Gamma_{DR}^{kd} \Gamma_{DL}^{ld*}) \right) \right] \frac{1}{\epsilon} \right. \\
& \left.
  + \frac{1}{3}
X_D^{kn} \, x_{d_n} \, X_D^{nl} \,  \frac{1}{\epsilon}  \right\} \quad
\mbox{ for } k\neq l \, .
\end{split}
\end{equation}
For $k=l$, we have $\delta\Gamma_{D}^{kl}=0$.
Detailed derivations of the renormalization of the squark and
gluino masses are given in 
Appendices~\ref{sec:squarkrenormalization} and \ref{sec:gluinorenormalization}, respectively.

Finally, we can put everything together to form the complete contributions of class b) in the full theory:
\begin{equation}\label{Arenbfull}
 \begin{split}
  i A_{b,full}^{ren} =& i A_{b,irred,full}^{bare} + i A_{b,red,full}^{bare}\\
 &+ 2\, \delta Z^b_{g_{s,Y}} \, \, \left(K^{(0)}_{7b,\tilde{g}} \,
\langle s \gamma |{\cal O}_{7b,\tilde{g}}|b\rangle_{\mbox{\tiny tree}} +  K^{(0)}_{7\tilde{g},\tilde{g}} \,
\langle s \gamma |{\cal O}_{7\tilde{g},\tilde{g}}|b\rangle_{\mbox{\tiny tree}} \right)  + R^b_{\tilde{g}\tilde{d}_k\Gamma} \, .
 \end{split}
\end{equation}

\subsubsection{Effective theory}
Since the contributions of class b) do not contain any gluon
corrections, the only one-loop contributions in the effective theory
are due to the box-operators in \eqref{boxop} and the box-induced part of
the operators in \eqref{penguinboxop} (given in expression~(32) in
\cite{Borzumati:1999qt}). 
The renormalized amplitude $iA_{b,eff}^{ren}$ can be written as
\begin{equation}
\begin{split}
         \label{Abrenaeff}
i \, A^{ren}_{b,eff} = & \frac{\alpha_s}{4\pi} \, C_{7g,g}^{(1),b} \,
\langle s \gamma |{\cal O}_{7\tilde{g},\tilde{g}}|b\rangle_{\mbox{\tiny tree}}
+ \frac{\alpha_s}{4\pi} \, C_{7b,g}^{(1),b} \,
\langle s \gamma |{\cal O}_{7b,\tilde{g}}|b\rangle_{\mbox{\tiny tree}}  + \frac{\alpha_s}{4\pi} \, C_{7c,g}^{(1),b} \,
\langle s \gamma |{\cal O}_{7c,\tilde{g}}|b\rangle_{\mbox{\tiny tree}}  \\
&+\sum\limits_{i=11}^{14} \sum\limits_q C_{i,\tilde{g}}^{(0),q,b}\, \langle s \gamma |{\cal O}^q_{i,\tilde{g}}|b\rangle_{\mbox{\tiny 1-loop}}  +\sum\limits_{i=15}^{20} \sum\limits_q C_{i,\tilde{g}}^{(0),q}\, \langle s \gamma |{\cal O}^q_{i,\tilde{g}}|b\rangle_{\mbox{\tiny 1-loop}} \\
& +\left(C_{15,\tilde{g}}^{(0),b} \delta Z_{15,7b} +C_{16,\tilde{g}}^{(0),b} \delta Z_{16,7b} \right)\langle s \gamma |{\cal O}_{7b,\tilde{g}}|b\rangle_{\mbox{\tiny tree}}\\
& +\left(C_{19,\tilde{g}}^{(0),b} \delta Z_{19,7b} +C_{20,\tilde{g}}^{(0),b} \delta Z_{20,7b} \right)\langle s \gamma |{\cal O}_{7b,\tilde{g}}|b\rangle_{\mbox{\tiny tree}}\\
& +\left(C_{19,\tilde{g}}^{(0),c} \delta Z_{19,7c} +C_{20,\tilde{g}}^{(0),c} \delta Z_{20,7c} \right)\langle s \gamma |{\cal O}_{7c,\tilde{g}}|b\rangle_{\mbox{\tiny tree}},
\end{split}
\end{equation}
with
\begin{equation}
\begin{aligned}
\delta Z_{15,7b} &= -\frac{\alpha_s}{4\pi} \frac{1}{\epsilon} \frac{1}{6}, \quad
 &\delta Z_{16,7b} = -\frac{\alpha_s}{4\pi} \frac{1}{\epsilon}\frac{1}{2},\\
 \delta Z_{19,7b} &= -\frac{\alpha_s}{4\pi} \frac{1}{\epsilon} \frac{1}{2},
 &\delta Z_{20,7b} = -\frac{\alpha_s}{4\pi} \frac{1}{\epsilon}\frac{1}{6},\\
 \delta Z_{19,7c} &= \frac{\alpha_s}{4\pi} \frac{1}{\epsilon} ,
 &\delta Z_{20,7c} = \frac{\alpha_s}{4\pi} \frac{1}{\epsilon} \frac{1}{3}.
 \end{aligned}
 \end{equation}

\subsubsection{Extracting $C^{(1),b}_{7\tilde{g},\tilde{g}}$, $C^{(1),b}_{7b,\tilde{g}}$ and  $C^{(1),b}_{7c,\tilde{g}}$ }
From the requirement that  \eqref{Arenbfull} and \eqref{Abrenaeff}  produce the same results, we can extract the next-to-leading order Wilson coefficients $C^{(1),b}_{7\tilde{g},\tilde{g}}$, $C^{(1),b}_{7b,\tilde{g}}$ and  $C^{(1),b}_{7c,\tilde{g}}$. Unfortunately, the results are much too long to be printed. We therefore provide a computer code (see Section~\ref{sec:results}), which enables the user to evaluate our results numerically for arbitrary input parameters.

\subsection{Results for $C^{(1)}_{7\tilde{g},\tilde{g}}$,
  $C^{(1)}_{7b,\tilde{g}}$, and  $C^{(1)}_{7c,\tilde{g}}$} 
We are now ready to recombine the contributions to the Wilson
coefficients of both  the gluon corrections of class a) and the gluino
corrections of class b) as 
\begin{equation}
 \begin{split}
  C^{(1)}_{7\tilde{g},\tilde{g}} &= C^{(1),a}_{7\tilde{g},\tilde{g}} +
  C^{(1),b}_{7\tilde{g},\tilde{g}},\\ 
  C^{(1)}_{7b,\tilde{g}} &= C^{(1),a}_{7b,\tilde{g}} + C^{(1),b}_{7b,\tilde{g}},\\
  C^{(1)}_{7c,\tilde{g}} &= C^{(1),a}_{7c,\tilde{g}} + C^{(1),b}_{7c,\tilde{g}}.\\
 \end{split}
\end{equation}
These results were, however, obtained by using dimensional
regularization (DREG) throughout the calculation followed by renormalizing
$m_{\tilde{g}}$, $m^2_{\tilde{d}_k}$, $g_{s,Y}$  and
$\Gamma_{DL/R}$ in the 
$\overline{\mbox{MS}}$ scheme. 
But, since DREG introduces a mismatch between the bosonic and
fermionic degrees of freedom, the present calculation in the full
theory has to be worked out in dimensional reduction (DRED)
\cite{Siegel:1979wq,Capper:1979ns}, which is expected to preserve
supersymmetry. In principle this requires either the calculation of
additional graphs featuring $\epsilon$-scalars or working out the
Dirac algebra in $d=4$ dimensions. 

At leading order in $\alpha_s$, the results in DREG and DRED are identical, while
at next-to-leading order the difference between DREG and DRED can be
obtained by shifting the strong coupling $g_s$ and the gluino mass
$m_{\tilde{g}}$ in the leading order contribution. Exploiting this, we
chose in a first step to evaluate all diagrams in the full theory
strictly in DREG, followed by renormalizing $m_{\tilde{d}_k}$,
$m_{\tilde{g}}$ and $g_s$ in the $\overline{\mbox{MS}}$ scheme. In
Section~\ref{sec:DRbarTrans}, we will discuss the shifts in $g_s$
and $m_{\tilde{g}}$.

Additionally, the calculation in the full theory was performed without
taking into account the decoupling of the heavy particles in the
running of the strong coupling. This will
be done in Section~\ref{sec:Decoupling}. 

\subsection{Transition to the $\overline{\mbox{DR}}$ scheme}\label{sec:DRbarTrans}

Since supersymmetry is broken in DREG, we have to distinguish between
the Yukawa-type coupling $g_{s,Y}$ appearing in the squark-quark-gluino
vertex and the gauge coupling $g_{s,G}$. Only supersymmetry guarantees
that both versions of the strong coupling are equal, which is the case
 in DRED accompanied by minimal subtraction (which is usually called 
$\overline{\mbox{DR}}$ scheme).

The result for the NLO full theory contributions obtained in the
$\overline{\mbox{MS}}$ scheme is expressed in terms of 
 $g_{s,Y}^{\overline{\mbox{\tiny MS}}}$ and
$m_{\tilde{g}}^{\overline{\mbox{\tiny MS}}}$. The transition to the
expression depending on the corresponding $\overline{\mbox{DR}}$
quantities amounts to performing the following replacements in the
leading order contributions
\cite{Martin:1993yx,Mihaila:2009bn}:\footnote{Note that $g_{s,G}$ does
not appear in our LO expressions.}
\begin{equation}\label{DRshift}
 \begin{split}
  g_{s,Y}^{\overline{\mbox{\tiny MS}}} &= 
  g_s^{\overline{\mbox{\tiny DR}}} 
  \left[1  + \frac{\alpha_s^{\overline{\mbox{\tiny DR}}}}{4 \pi}
  \frac{(C_A-C_F)}{2}\right],\\
  m_{\tilde{g}}^{\overline{\mbox{\tiny MS}}} &=
  m_{\tilde{g}}^{\overline{\mbox{\tiny DR}}}\left[ 1+
    \frac{\alpha_s^{\overline{\mbox{\tiny DR}}}}{4 \pi} C_A \right]. 
 \end{split}
\end{equation}
This leads to finite shifts of the NLO Wilson coefficients which we
denote by $\delta C_{7\tilde{g},\tilde{g}}^{(1),\overline{\mbox{\tiny
DR}}}$ and $\delta C_{7b,\tilde{g}}^{(1),\overline{\mbox{\tiny
DR}}}$. 
Since we treat the external $b$ and $s$ quarks in the on-shell scheme, 
their masses do not need to be shifted. 

\subsection{Decoupling of heavy particles}\label{sec:Decoupling}

After performing the shifts in \eqref{DRshift}, the result on the full
theory side depends on 
$\alpha_{s,\overline{\mbox{\tiny DR}}}^{\mbox{\tiny (full)}}$.  
However, on the effective side, there are only five active quark
flavors and the coupling is renormalized in the
$\overline{\mbox{MS}}$ scheme. Hence, we have to express
$\alpha_{s,\overline{\mbox{\tiny DR}}}^{\mbox{\tiny (full)}}$ in terms
of $\alpha_{s,\overline{\mbox{\tiny MS}}}^{(5)}$ on the full theory
side. The relation between $\alpha_{s,\overline{\mbox{\tiny
      DR}}}^{\mbox{\tiny (full)}}$ and $\alpha_{s,\overline{\mbox{\tiny
      DR}}}^{(5)}$ can be found in \cite{Harlander:2005wm} as 
\begin{equation}\label{eqn:decoupling1}
  \alpha_{s,\overline{\mbox{\tiny DR}}}^{\mbox{\tiny (full)}} =
  \left(\zeta^{\mbox{\tiny
      SUSY}}_{g_s}\right)^{-2}\alpha_{s,\overline{\mbox{\tiny
        DR}}}^{(5)}, 
\end{equation}
with the decoupling coefficient
\begin{equation}
 \left(\zeta^{\mbox{\tiny SUSY}}_{g_s}\right)^{-2} = 1 -
 \frac{\alpha_{s,\overline{\mbox{\tiny DR}}}^{(5)}}{\pi} \left[
   -\frac{1}{6}
   \ln\frac{\mu^2}{m_t^2}-\frac{1}{24}\left(\sum\limits_{k=1}^6
   \ln\frac{\mu^2}{m^2_{\tilde{u}_k}} + \sum\limits_{k=1}^6
   \ln\frac{\mu^2}{m^2_{\tilde{d}_k}}\right)-\frac{1}{2}
   \ln\frac{\mu^2}{m^2_{\tilde{g}}}\right]. 
\end{equation}
We now express  $\alpha_{s,\overline{\mbox{\tiny DR}}}^{(5)}$ in terms
of  $\alpha_{s,\overline{\mbox{\tiny MS}}}^{(5)}$ according to 
\begin{equation}\label{eqn:decoupling2}
  \alpha_{s,\overline{\mbox{\tiny DR}}}^{(5)} =
  \alpha_{s,\overline{\mbox{\tiny MS}}}^{(5)}
  \left[1+\frac{\alpha_{s,\overline{\mbox{\tiny
            MS}}}^{(5)}}{4\pi}\frac{C_A}{3}\right]. 
\end{equation}
Combining \eqref{eqn:decoupling1} and \eqref{eqn:decoupling2} we can
finally give the relation between $\alpha_{s,\overline{\mbox{\tiny
      DR}}}^{\mbox{\tiny (full)}}$ and
$\alpha_{s,\overline{\mbox{\tiny MS}}}^{(5)}$ as 
\begin{equation}
 \alpha_{s,\overline{\mbox{\tiny DR}}}^{\mbox{\tiny (full)}} =
 \alpha_{s,\overline{\mbox{\tiny MS}}}^{(5)} \left\{1  +
 \frac{\alpha_{s,\overline{\mbox{\tiny MS}}}^{(5)}}{\pi} \left[\frac{
     C_A}{12} +\frac{1}{6}
   \ln\frac{\mu^2}{m_t^2}+\frac{1}{24}\left(\sum\limits_{k=1}^6
   \ln\frac{\mu^2}{m^2_{\tilde{u}_k}} + \sum\limits_{k=1}^6
   \ln\frac{\mu^2}{m^2_{\tilde{d}_k}}\right)+\frac{1}{2}
   \ln\frac{\mu^2}{m^2_{\tilde{g}}}\right]\right\}. 
\end{equation}
This replacement leads to a finite shift of the NLO Wilson
coefficients which we denote by $\delta
C_{7\tilde{g},\tilde{g}}^{(1),dec.}$ and $\delta
C_{7b,\tilde{g}}^{(1),dec.}$.

\subsection{Final result for $C_{7\tilde{g},\tilde{g}}^{(1)}$, $C_{7b,\tilde{g}}^{(1)}$, and $C_{7c,\tilde{g}}^{(1)}$}
We now have all ingredients to give the final result for all NLO Wilson coefficients of the magnetic operators. They read
\begin{equation}
 \begin{split}
  C^{(1),\overline{\mbox{\tiny DR}}}_{7\tilde{g},\tilde{g}} =& C_{7\tilde{g},\tilde{g}}^{(1),a} +  C_{7\tilde{g},\tilde{g}}^{(1),b} + \delta C_{7\tilde{g},\tilde{g}}^{(1),\overline{\mbox{\tiny DR}}} +\delta C_{7\tilde{g},\tilde{g}}^{(1),dec.},\\
  C^{(1),\overline{\mbox{\tiny DR}}}_{7b,\tilde{g}} =& C_{7b,\tilde{g}}^{(1),a} +  C_{7b,\tilde{g}}^{(1),b} + \delta C_{7b,\tilde{g}}^{(1),\overline{\mbox{\tiny DR}}} +\delta C_{7b,\tilde{g}}^{(1),dec.},\\
  C^{(1),\overline{\mbox{\tiny DR}}}_{7c,\tilde{g}} =& C_{7c,\tilde{g}}^{(1),a} +  C_{7c,\tilde{g}}^{(1),b}.
 \end{split}
\end{equation}

\section{Calculation of the Wilson coefficients of the\\ chromomagnetic operators}\label{sec:chromo}
In order to extract the NLO Wilson coefficients of the chromomagnetic operators, we have to match the full theory NLO result for the amplitude of $b\to s g$ to the one in the effective theory. Since the Dirac structure of the photon vertices is the same as the one of the gluon vertices, we can use the results for the diagrams already calculated for $b\to s\gamma$ and supplement them with the appropriate color factor. In addition to these diagrams, we have to take into account all those where the gluon couples to gluino and gluon lines.

In full analogy  to the procedure for extracting the Wilson coefficients of the magnetic operators, we also divide the calculation into the classes a) and b) introduced in Section~\ref{sec:magnetic}.

\subsection{Results for $C_{8\tilde{g},\tilde{g}}^{(1),a}$ and $C_{8b,\tilde{g}}^{(1),a}$}
Since the calculation for the gluon corrections has been presented in
\cite{Bobeth:1999ww} and we cross-checked the results for $b\to
s\gamma$, we do not  repeat the calculation and directly take the
result from \cite{Bobeth:1999ww}. It reads
\begin{equation}
\begin{split}
\label{resultpart18}
& C^{(1),a}_{8\tilde{g},\tilde{g}} =
\frac{4}{3} \, \frac{1}{16\pi^2} \, \frac{1}{m_{\tilde{g}}} \, 
\sum\limits_{k=1}^6 \Gamma_{DR}^{kb} \Gamma_{DL}^{ks*} \, 
 h^{(1),a}_{8\tilde{g},\tilde{g}}(x_{d_k})  ,\\
& C^{(1),a}_{8b,\tilde{g}} =
- \frac{4}{3} \, \frac{1}{16\pi^2} \, \frac{1}{m_{\tilde{g}}^2} \, 
\sum\limits_{k=1}^6 \Gamma_{DL}^{kb} \Gamma_{DL}^{ks*} \, 
h^{(1),a}_{8b,\tilde{g}}(x_{d_k}) ,
\end{split}
\end{equation}
with
\begin{equation}
 \begin{split}
  h^{(1),a}_{8\tilde{g},\tilde{g}}(x)=&\left(\frac{7 \left(79 x^2-12 x+5\right)}{12 (x-1)^3}-\frac{7 x \left(18 x^2+19 x-1\right) }{6
   (x-1)^4}\ln (x)\right)L_\mu\\
   &+\frac{\left(-359 x^2+339 x-204\right)}{24 (x-1)^3} \text{Li}_2(1-x)+\frac{x \left(181 x^2-236 x+31\right) }{48 (x-1)^4}\ln ^2(x)\\
   &-\frac{\left(2419 x^3-771 x^2+453 x+11\right)
   }{48 (x-1)^4}\ln (x)+\frac{1667 x^2-990 x+379}{24 (x-1)^3},\\[2mm]
     h^{(1),a}_{8b,\tilde{g}}(x)=& \left(\frac{x \left(747 x^2+640 x-43\right) }{48 (x-1)^5}\ln (x)-\frac{779 x^3+7203 x^2+93 x-11}{288
   (x-1)^4}\right) L_\mu\\
   &+\frac{\left(-45 x^3+1208 x^2-901 x+570\right) }{96 (x-1)^4}\text{Li}_2(1-x)-\frac{x \left(45 x^2+358
   x-49\right) }{192 (x-1)^4}\ln ^2(x)\\
   &+\frac{\left(-183
   x^4+30027 x^3-10692 x^2+6115 x+77\right) }{864 (x-1)^5}\ln (x)\\
   &+\frac{5359 x^3-241425 x^2+143253 x-59251}{5184 (x-1)^4}.
 \end{split}
\end{equation}

\subsection{Calculation of $C_{8\tilde{g},\tilde{g}}^{(1),b}$, $C_{8b,\tilde{g}}^{(1),b}$, and $C_{8c,\tilde{g}}^{(1),b}$}
Apart from additional diagrams, there are two new features in this calculation on the full theory side: Firstly, the emitted gluon couples with the gauge version of the strong coupling $g_{s,G}$, which, at leading order, is different from the Yukawa version. Secondly, the gluon field has to be renormalized. The required  renormalization constants are derived in Appendices~\ref{gGren} and \ref{sec:gluinorenormalization}, respectively. They read
\begin{equation}
 \begin{split}
  \delta Z_{g_s,G}^b &= \frac{\alpha_s}{4\pi} 
\left[ \frac{1}{3} \, tr \, n_f + \frac{C_A}{3} \right]\frac{1}{\epsilon} ,\\[1mm]
\delta Z_3^b &= -2\frac{\alpha_s}{4\pi} \left[ \frac{1}{3} \, tr \, n_f + \frac{C_A}{3} \right]
\frac{1}{\epsilon}.
 \end{split}
\end{equation}
Multiplying the leading order contribution on the full theory side, they both induce counter\-terms at the next-to-leading order. However, they  appear in the combination
\begin{equation}
 \delta Z_{g_s,G}^b + \frac{1}{2}\delta Z_3^b = 0,
\end{equation}
and therefore their effect cancels.

The calculation of the decay amplitude for $b\to s g$ on the effective side is very similar to the one for the $b\to s \gamma$ transition. The non-vanishing renormalization constants for the operators read
\begin{eqnarray}
\delta Z_{15,8b} = \frac{\alpha_s}{4\pi} \frac{1}{\epsilon}\frac{1}{2}, \qquad
\delta Z_{20,8b} = \frac{\alpha_s}{4\pi} \frac{1}{\epsilon}\frac{1}{2}, \qquad
 \delta Z_{20,8c} = \frac{\alpha_s}{4\pi} \frac{1}{\epsilon} \frac{1}{2}.
\end{eqnarray}

After the extraction of the Wilson coefficients, the results are expressed in $\overline{\mbox{MS}}$ entities and we need to replace them by their $\overline{\mbox{DR}}$ counterparts. In contrast to the $b\to s\gamma$ case, the leading order contribution now also contains the gauge coupling $g_{s,g}^{\overline{\mbox{\tiny MS}}}$ which has to be replaced according to \cite{Mihaila:2009bn}
\begin{equation}
 g_{s,G}^{\overline{\mbox{\tiny MS}}} =g_s^{\overline{\mbox{\tiny DR}}} \left[1-\frac{\alpha_s^{\overline{\mbox{\tiny DR}}}}{4 \pi} \frac{C_A}{6}\right].
\end{equation}

The decoupling of the heavy particles and the shift back to the $\overline{\mbox{MS}}$ scheme works exactly the same way as for $b\to s\gamma$.

\section{Results for the Wilson coefficients at the matching scale $\mu_W$}\label{sec:results}
As mentioned above, the analytic results for the NLO Wilson coefficients are too long to be printed. We therefore provide a C++ code available at {\ttfamily  arXiv.org}  which enable the user to evaluate our results numerically for arbitrary input parameters. We also include the results for the LO Wilson coefficients that were given in \cite{Borzumati:1999qt}.

The file {\ttfamily  bsg2gluino.h} contains the definitions of the
functions for all LO and NLO Wilson coefficients and can be included
into any C++ code.\footnote{An equivalent FORTRAN version is available
  from C.G.} 
All input parameters, namely the rotation matrices
of the squark fields $\Gamma_{(U,D)(L,R)}$ defined in
(\ref{squark_rotation}), the squark masses, the gluino mass
$m_{\tilde{g}}$, the top mass $m_t$, the bottom mass $m_b$, the strange mass $m_s$, and the down mass $m_d$, are declared as global variables within {\ttfamily  bsg2gluino.h} according to the naming scheme shown in Table~\ref{cppParams}.
\begin{table}
\renewcommand{\arraystretch}{1.5}
\begin{center}
\begin{tabular}{|ll|}
\hline
Parameter & C++ variable\\ \hline
\rule{0cm}{4ex} 
$(\Gamma_{UL},\Gamma_{UR})$ & 
{\ttfamily std::complex<double> GU[6][6]}\\
$(\Gamma_{DL},\Gamma_{DR})$ & 
{\ttfamily std::complex<double> GD[6][6]}\\
$(m_{\tilde{u_1}}^2,\ldots, m_{\tilde{u_6}}^2)$ &
{\ttfamily double mupsquark[6]}\\
$(m_{\tilde{d_1}}^2,\ldots, m_{\tilde{d_6}}^2)$&
{\ttfamily double mdownsquark[6]}\\
$m_{\tilde{g}}$ & {\ttfamily double mg}\\
$m_t$ & {\ttfamily double mt}\\
$m_b$ & {\ttfamily double mb}\\
$m_s$ & {\ttfamily double ms}\\
$m_d$ & {\ttfamily double md}\\\hline
\end{tabular}
\caption{Input parameters.}
\label{cppParams}
\end{center}
\end{table}

The names for the provided functions for the Wilson coefficients are listed in
Tables~\ref{cppLOWC} and~\ref{cppNLOWC}. We define the expansion
in $\alpha_s$ as usual by
\begin{equation}
C_i = C_i^{(0)}+\frac{\alpha_s}{4\pi}C_i^{(1)}.
\end{equation}
The NLO Wilson coefficients explicitly depend on the
matching scale $\mu_W$. Hence, these functions have to be provided
with a value for $\mu_W$ as an argument. 

To illustrate the use of {\ttfamily  bsg2gluino.h}, we also provide
the compilable file {\ttfamily  template.cc} in which the header
{\ttfamily  bsg2gluino.h} is included and all input parameters are set
to some specific values. The compiled program will then generate
output for all included Wilson coefficients as shown in the file {\ttfamily output.dat}. 

\begin{table}
\renewcommand{\arraystretch}{1.5}
\begin{center}
\begin{tabular}{|ll|}
\hline
Wilson coefficient & C++ function \\ \hline
\rule{0cm}{4ex} 
$C_{7\tilde{g}\tilde{g}}^{(0)}$ & 
{\ttfamily std::complex<double> C7ggLO()}\\
$C_{7\tilde{g}\tilde{g}}^{\prime(0)}$ & 
{\ttfamily std::complex<double> C7ggLOPrime()}\\
$C_{7b\tilde{g}}^{(0)}$ & 
{\ttfamily std::complex<double> C7bgLO()}\\
$C_{7b\tilde{g}}^{\prime(0)}$ & 
{\ttfamily std::complex<double> C7bgLOPrime()}\\
$C_{8\tilde{g}\tilde{g}}^{(0)}$ & 
{\ttfamily std::complex<double> C8ggLO()}\\
$C_{8\tilde{g}\tilde{g}}^{\prime(0)}$ & 
{\ttfamily std::complex<double> C8ggLOPrime()}\\
$C_{8b\tilde{g}}^{(0)}$ & 
{\ttfamily std::complex<double> C8bgLO()}\\
$C_{8b\tilde{g}}^{\prime(0)}$ & 
{\ttfamily std::complex<double> C8bgLOPrime()}\\
$C_{i\tilde{g}}^{q,(0)}$& 
{\ttfamily std::complex<double> C`i'g`q'LO()}, e.g. 
{\ttfamily C11guLO()}\\
$C_{i\tilde{g}}^{q,\prime (0)}$& 
{\ttfamily std::complex<double> C`i'g`q'LOPrime()}\\
\hline
\end{tabular}
\end{center}
\caption{C++ functions for the LO Wilson coefficients with $i=11,\ldots,20$ and $q=u,d,c,s,b$.}
\label{cppLOWC}
\end{table}

\begin{table}
\renewcommand{\arraystretch}{1.5}
\begin{center}
\begin{tabular}{|ll|}
\hline
Wilson coefficient & C++ function \\ \hline
\rule{0cm}{4ex} 
$C_{i\tilde{g}\tilde{g}}^{\overline{\text{DR}},(1)}(\mu)$ & 
{\ttfamily std::complex<double> C`i'ggDRbar(double)}\\
${C_{i\tilde{g}\tilde{g}}^{\overline{\text{DR}},(1)}}^\prime(\mu)$ & 
{\ttfamily std::complex<double> C`i'ggDRbarPrime(double)}\\
$C_{ib\tilde{g}}^{\overline{\text{DR}},(1)}(\mu)$ & 
{\ttfamily std::complex<double> C`i'bgDRbar(double)}\\
${C_{ib\tilde{g}}^{\overline{\text{DR}},(1)}}^\prime(\mu)$ & 
{\ttfamily std::complex<double> C`i'bgDRbarPrime(double)}\\
$C_{ic\tilde{g}}^{\overline{\text{DR}},(1)}(\mu)$ & 
{\ttfamily std::complex<double> C`i'cgDRbar(double)}\\
${C_{ic\tilde{g}}^{\overline{\text{DR}},(1)}}^\prime(\mu)$ & 
{\ttfamily std::complex<double> C`i'cgDRbarPrime(double)}\\
\hline
\end{tabular}
\end{center}
\caption{C++ functions for the NLO Wilson coefficients with $i=7,8$.}
\label{cppNLOWC}
\end{table}

\section{\label{sec:concl}Conclusion}
The main purpose of this paper is the two-loop matching calculation of
the gluino-induced contributions to the processes $b\to s \gamma$ and
$b \to s g$. While the contribution involving one gluino has already
been available in the literature, the two-gluino part is new. We
discuss in detail our renormalization procedure and the issues related to dimensional reduction and the decoupling of heavy
particles from the running of $\alpha_s$. Our results are presented in
the C++ file {\tt bsg2gluino.h} which comes together with this paper
and allows for a convenient numerical evaluation.

The results obtained in this paper constitute a crucial building block for the
phenomenological NLL  analysis of the branching ratio $\bar{B}\to X_s
\gamma$ in a supersymmetric model beyond MFV because 
the NLL gluino  contributions are the most important
corrections beyond MFV at this order in most parts of the parameter space in such models.

\section{Acknowledgements}

C.G. and C.S. were partially supported by the Swiss National Foundation and by
the Helmholz Association through 
funds provided to the virtual institute ``Spin and strong QCD''
(VH-VI-231). The Albert Einstein Center for Fundamental Physics
(Bern), to which C.G. and C.S. are affiliated, is supported by the
``Innovations- und Kooperationsprojekt C-13 of the Schweizerische Universit\"atskonferenz SUK/CRUS''.
M.S. is partially supported by the DFG through the SFB/TR~9 ``Computational
Particle Physics''.
T.H. thanks the CERN theory group for its hospitality during his regular vists to CERN.
Finally, we all thank Andreas Crivellin for very helpful and
interesting discussions.

\clearpage

\begin{appendix}

\section*{\huge Appendix}

In Appendix \ref{sec:LOresult} we give the LO expressions for the Wilson
coefficients (in $d$ dimensions) which are used at various places in this paper.
Appendix \ref{sec:quarkrenormalization} is devoted to the evaluation of the renormalization
constants of the quark fields and quark masses in the on-shell scheme.
Appendices  \ref{sec:squarkrenormalization} to \ref{sec:gluonrenormalization}
deal with the corresponding renormalization constants related to the squarks,
the gluino and the gluon. For completeness we give all these quantities in the
on-shell scheme, although for some of them only the singular piece (in
$\epsilon$) is needed in the main text. 
Finally, in Appendix \ref{renormg} we derive the $1/\epsilon$ pieces of
the renormalization constants for the strong coupling constants
$g_{s,Y}$ and $g_{s,G}$, as well as for the $\Gamma_{DL/R}$ matrices. 
All the results in the Appendix are obtained by using DREG.

 \section{Leading order result for $b \to s \gamma$ in the full theory}\label{sec:LOresult}
Using the abbreviations
$x_{d_k} =m_{\tilde{d}_k}^2/m_{\tilde{g}}^2$ and
$L_\mu=\ln(\mu^2/m_{\tilde{g}}^2)$,
the decay amplitude  ${\cal A}(b \to s \gamma)$ 
in the full theory to leading
order in QCD 
(corresponding to the diagrams with one-loop gluino/squark
exchange) can be written for a massless $s$ quark as 
\begin{equation}
\begin{split}
\label{absgammaleading}
i A_{full}(b \to s \gamma) =&  K^{(0)}_{7b,\tilde{g}} \, \langle s \gamma|{\cal
  O}_{7b,\tilde{g}}|b \rangle_{\mbox{\tiny tree}} 
+ K^{(0)}_{7\tilde{g},\tilde{g}} \, \langle s \gamma|{\cal
  O}_{7\tilde{g},\tilde{g}}|b \rangle_{\mbox{\tiny tree}}  \\
& + K^{(0)'}_{7b,\tilde{g}} \, \langle s \gamma|{\cal
  O}_{7b,\tilde{g}}^{'}|b \rangle_{\mbox{\tiny tree}} 
+ K^{(0)'}_{7\tilde{g},\tilde{g}} \, \langle s \gamma|{\cal
  O}_{7\tilde{g},\tilde{g}}^{'}|b \rangle_{\mbox{\tiny tree}} \, .
\end{split}
\end{equation}
The first two terms on the r.h.s. of (\ref{absgammaleading}) 
correspond to a left handed $s$ quark and the last two terms
to a right handed $s$ quark.
The various coefficients $K^{(0)}$ read (retaining terms of order~$\epsilon^1$)
\begin{equation}
 \begin{split}
 K^{(0)}_{7b,\tilde{g}} =& \sum\limits_{k=1}^6 \frac{C_F \Gamma_{DL}^{kb} \Gamma_{DL}^{ks*} Q_d}{16 \pi ^2 m_{\tilde{g}}^2 \left(x_{d_k}-1\right)^4}
  \Biggl\{\frac{1}{12} \left(-x_{d_k}^3+6 x_{d_k}^2-3 x_{d_k}-2\right)-\frac{1}{2}x_{d_k}\ln(x_{d_k})     \Biggr.\\
  &\left.+\epsilon \left[-\frac{\left(x_{d_k}-1\right)}{72}  \left(5 x_{d_k}^2-49 x_{d_k}-22+6 \left(x_{d_k}^2-5x_{d_k}-2\right) L_\mu\right)\right.\right.\\
  &\left.\biggl. +\frac{1}{12} x_{d_k} \ln (x_{d_k})
    \left(x_{d_k}^2-6 x_{d_k}-6-6L_\mu\right)+\frac{1}{4}x_{d_k} \ln
    ^2(x_{d_k})\biggr] 
\right\},
 \end{split}
\end{equation}

\begin{equation}
 \begin{split}
K^{(0)}_{7\tilde{g},\tilde{g}} =&\sum\limits_{k=1}^6 \frac{C_F \Gamma_{DR}^{kb} \Gamma_{DL}^{ks*} Q_d}{16 \pi ^2 m_{\tilde{g}} \left(x_{d_k}-1\right)^3} 
\left\{\frac{1}{2}  \left(x_{d_k}^2-1\right)-  x_{d_k} \ln (x_{d_k}) + \epsilon \left[\frac{1}{4} \left(x_{d_k}^2-1\right) \left(3+2
  L_\mu\right)  \right.  \right.\\
&\left.\left.- \frac{1}{2} x_{d_k} \ln (x_{d_k}) \left(x_{d_k}+2+2
  L_\mu\right)+\frac{1}{2}  x_{d_k} \ln ^2(x_{d_k})\right]  \right\}.
 \end{split}
\end{equation}
The corresponding primed coefficients $K^{(0)'}_{7b,\tilde{g}}$ and
$K^{(0)'}_{7\tilde{g},\tilde{g}}$ can be obtained by interchanging $L
\leftrightarrow R$ in the unprimed versions. Note that we explicitly kept terms 
of order $\epsilon^1$. For the individual equations
(\ref{abareafulls}), (\ref{Arenafull}) and (\ref{Arenaeff}) even the
$\epsilon^2$ contributions of these coefficients would be required.
In the matching equation (\ref{matchingequation}), they drop
out, however. 

The decay amplitude  ${\cal A}(b \to s g)$ in the full theory to leading
order in QCD 
(corresponding to the diagrams with one-loop gluino/squark
exchange) can be written in complete analogy to
 (\ref{absgammaleading}).
The corresponding coefficients read
\begin{equation}
 \begin{split}
 K^{(0)}_{8b,\tilde{g}} =&\sum\limits_{k=1}^6 \frac{\Gamma_{DL}^{kb}\Gamma_{DL}^{ks*}}{16 \pi ^2 m_{\tilde{g}}^2 \left(x_{d_k}-1\right)^4} 
 \left\{\left(C_F-\frac{1}{2}C_A\right) \left[\frac{1}{12} \left(-x_{d_k}^3+6 x_{d_k}^2-3 x_{d_k}-2\right) -\frac{1}{2}x_{d_k} \ln(x_{d_k})\right.\right.\\
 &\left.\left.+ \epsilon \left(\frac{1}{12} x_{d_k} \ln (x_{d_k}) \left(-6 L_\mu+x_{d_k}^2-6 x_{d_k}-6\right)+\frac{1}{4}
   x_{d_k} \ln ^2(x_{d_k})\right.\right.\right.\\
 &\left.\left.\left.-\frac{1}{72} \left(x_{d_k}-1\right) \left(6 \left(x_{d_k}^2-5
   x_{d_k}-2\right) L_\mu+5 x_{d_k}^2-49 x_{d_k}-22\right)\right) \right]\right.\\
   &\left.+\frac{1}{2} C_A \left[-\frac{1}{2} x_{d_k}^2 \ln (x_{d_k})+\frac{1}{12} \left(2 x_{d_k}^3+3 x_{d_k}^2-6 x_{d_k}+1\right)\right.\right.\\
   &\left.\left.+\epsilon \left(-\frac{1}{12} x_{d_k}^2 \ln (x_{d_k})\left(6 L_\mu+2 x_{d_k}+9\right)+\frac{1}{4}x_{d_k}^2 \ln ^2(x_{d_k})\right.\right.\right.\\
   &\left.\left.\left.+\frac{1}{72} \left(x_{d_k}-1\right) \left(6
   \left(2x_{d_k}^2+5 x_{d_k}-1\right) L_\mu+22 x_{d_k}^2+49
   x_{d_k}-5\right)\right) \right]\right\},
   \end{split}
\end{equation}
\begin{equation}
\begin{split}
 K^{(0)}_{8\tilde{g},\tilde{g}} =&
  \sum\limits_{k=1}^6\frac{\Gamma_{DR}^{kb}\Gamma_{DL}^{ks*}}{16 \pi ^2 m_{\tilde{g}} \left(x_{d_k}-1\right)^3} 
  \left\{\left (C_F-\frac{1}{2} C_A \right) \left[\frac{1}{2}  \left(x_{d_k}^2-1\right)-
   x_{d_k} \ln (x_{d_k})\right.\right.\\
  & \left.\left.+\epsilon \left(\frac{1}{4}  \left(x_{d_k}^2-1\right)
   \left(2L_\mu+3\right)-\frac{1}{2}  x_{d_k} \ln (x_{d_k}) \left(2
   L_\mu+x_{d_k}+2\right)+\frac{1}{2}  x_{d_k} \ln ^2(x_{d_k})\right)
   \right]\right.\\
   &\left.+\frac{1}{2} C_A  \left[-  x_{d_k}^2 \ln(x_{d_k})+\frac{1}{2}  \left(3 x_{d_k}^2-4 x_{d_k}+1\right)\right.\right.\\
   &\left.\left.+\epsilon \left(\frac{1}{2}  x_{d_k}^2 \ln ^2(x_{d_k})-\frac{1}{2} x_{d_k}^2 \ln (x_{d_k}) \left(2 L_\mu+3\right)  \phantom{\frac{1}{1}}\right.\right.\right.\\
   &\left.\left.\left.+\frac{1}{4}  \left(x_{d_k}-1\right)
    \left(\left(6 x_{d_k}-2\right) L_\mu+7 x_{d_k}-1\right)\right) \right]\right\}.
 \end{split}
\end{equation}
The corresponding primed coefficients $K^{(0)'}_{8b,\tilde{g}}$ and
$K^{(0)'}_{8\tilde{g},\tilde{g}}$ can again be obtained by interchanging $L
\leftrightarrow R$ in the unprimed versions.

\section{Quark-field and quark-mass renormalization in the on-shell scheme}
\label{sec:quarkrenormalization}
We start with the free (bare) quark Lagrangian for the down-type quarks
\begin{equation}\begin{split}
{\cal L} = i \, \bar{\psi}_{i,L}^0 \, \gamma^\mu \partial_\mu \delta_{ij}
\psi_{j,L}^0 
         + i \, \bar{\psi}_{i,R}^0 \, \gamma^\mu \partial_\mu \delta_{ij}
 \psi_{j,R}^0 - \bar{\psi}_{i,R}^0 m_i^0 \psi_{i,L}^0
- \bar{\psi}_{i,L}^0 (m_i^0)^* \psi_{i,R}^0 \,,
 \end{split}\end{equation}  
and express the bare quantities (marked with the superscript $(0)$) in
terms of renormalized ones:
\begin{equation}\begin{split}
& \psi_{i,L}^0 = \left( \delta_{ij} + \frac{1}{2} \delta  {
  Z}^{L}_{ij} \right) \, \psi_{j,L} \, , \qquad
\psi_{i,R}^0 = \left( \delta_{ij} + \frac{1}{2} \delta  {
  Z}^{R}_{ij} \right) \, \psi_{j,R} \, ,  \\
& m_{i}^0 = m_{i} - \delta m_i \, .
\end{split}\end{equation}
For the renormalized quark self-energies $\Sigma_{ji}^{ren}$ we get
\begin{equation}\begin{split}
\Sigma_{ji}^{ren} =& \Sigma_{ji} - \frac{1}{2} (\delta Z^L +
\delta Z^{L\dagger})_{ji} \, \pslash \, P_L  
 - \frac{1}{2} (\delta Z^R + \delta Z^{R \dagger})_{ji}
 \, \pslash \, P_R 
 \\
 & + \frac{1}{2} (\delta Z^{R \dagger})_{ji} \, m_i \, P_L +  
      \frac{1}{2} (\delta Z^{L \dagger})_{ji} \, m_i \, P_R +
     \frac{1}{2}  \, m_j \, \delta Z^L_{ji} \, P_L +  
      \frac{1}{2} \, m_j \, \delta Z^R_{ji} \, P_R  \\
& - \delta m_i \, \delta_{ji} \, P_L     
   - \delta m_i^* \, \delta_{ji} \, P_R \, ,      
\end{split}\end{equation}
where $\Sigma_{ji}$ denotes the bare quark self-energies which 
can be decomposed as  
\begin{equation}
\Sigma_{ji} = \pslash \, P_L \, \Sigma_{ji}^L(p^2) + \pslash \, P_R \,
\Sigma_{ji}^R(p^2) + P_L \, A_{ji}^L(p^2)+ P_R \, A_{ji}^R(p^2) \, .
\end{equation}
\begin{figure}
 \subfigure[]{
  \includegraphics[scale=.7, trim= 0 -5.5 0 0]{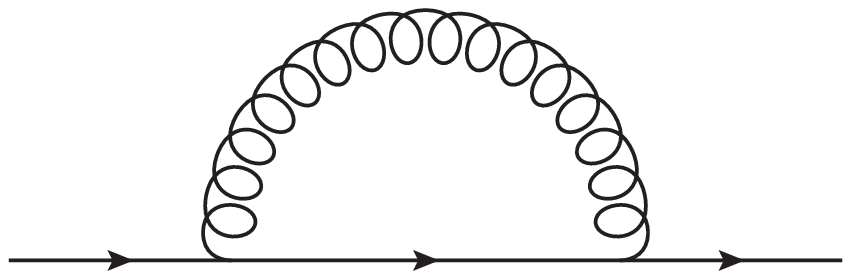}
  }
 \subfigure[]{
  \includegraphics[scale=.7]{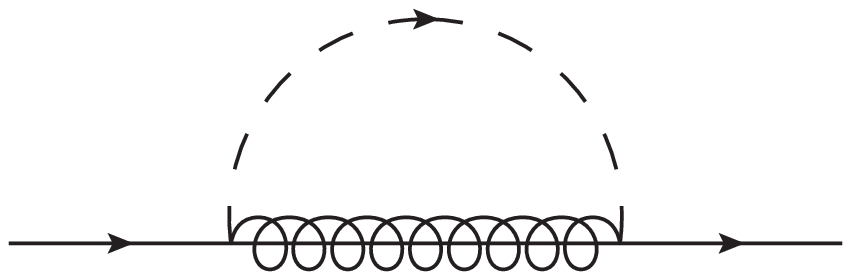}
  }\caption{Contributions to the bare quark self-energies $\Sigma_{ji}$.}
\label{Fig:quarkself}
 \end{figure}
We note that for $j \neq i$ the bare self-energies at order $\alpha_s^1$
receive only
contributions from diagram (b) in Figure \ref{Fig:quarkself} 
with a squark and a gluino in the loop,
while in the diagonal case $j=i$ also diagram (a) 
with an internal quark and a gluon contributes. 

\subsection{Off-diagonal case}
For $j \neq i$ there are only contributions from the gluino-squark loops shown in Figure~\ref{Fig:quarkself} (b). Because of this, we supplement the renormalization constants with the label $b$.
The on-shell renormalization conditions read
\begin{equation}\begin{split}
    &\left.\widetilde{\text{Re}}\,\Sigma_{ji}^{ren} \, u(p_i) \stackrel{!}{=} 0 \right|_{p_i^2=m_i^2,} \\
    &\left.\widetilde{\text{Re}}\,\bar{u}(p_j)\Sigma_{ji}^{ren}
    \stackrel{!}{=} 0\right|_{p_j^2=m_j^2}
    \,,
\end{split}\end{equation}
where the symbol $\widetilde{\text{Re}}$ stands for taking the dispersive part.
From these conditions, the wave function renormalization constants can
be fixed uniquely. The generic formulas
for the hermitian and 
antihermitian parts of $\delta Z_{ji}^{b,L}$ read 
(written in terms of
$\Sigma_{ji}^L$, $\Sigma_{ji}^R$, $A_{ji}^L$, and $A_{ji}^R$)
\begin{equation}
\begin{split}
\delta Z_{ji}^{b,L,H}=& \frac{1}{m_i^2-m_j^2}\widetilde{\text{Re}}\left\{ m_i [A^R_{ji}(m_i^2)- A^R_{ji}(m_j^2)]+m_j[A^L_{ji}(m_i^2)- A^L_{ji}(m_j^2)]\right.\\
&\left.+m_i m_j[\Sigma^R_{ji}(m_i^2)- \Sigma^R_{ji}(m_j^2)]+m_i^2 \Sigma^L_{ji}(m_i^2)-m_j^2\Sigma^L_{ji}(m_j^2)\right\},\\
    \delta Z_{ji}^{b,L,AH}=& \frac{1}{m_i^2-m_j^2}\widetilde{\text{Re}}\left\{m_i [A^R_{ji}(m_j^2)+ A^R_{ji}(m_i^2)]+m_j [A^L_{ji}(m_i^2)+ A^L_{ji}(m_j^2)]\right.\\
   &\left.+m_i m_j [\Sigma^R_{ji}(m_i^2)+
      \Sigma^R_{ji}(m_j^2)]+m_i^2\Sigma^L_{ji}(m_i^2)+m_j^2
    \Sigma^L_{ji}(m_j^2)\right\} \, . \\
\end{split}
\end{equation}
The right handed versions can be obtained by replacing all labels
$R\leftrightarrow L$. Due to the unitarity of the $\Gamma$ matrices,
the off-diagonal renormalization constants turn out to be finite
  for $d=4$. They are needed to order $\epsilon^0$ and up to linear
order in the quark masses $m_i$ and $m_j$. Explicitly, we find
\begin{equation}
  \begin{split}
    \delta Z_{ji}^{b,L,AH}=& \frac{\alpha_s}{4 \pi}
    \frac{C_F}{m_i^2-m_j^2} \sum\limits_{k=1}^6
    \Biggl\{(\Gamma_{DL}^{ki} \Gamma_{DR}^{kj*} m_j+\Gamma_{DL}^{kj*}
    \Gamma_{DR}^{ki} m_i) \left(\frac{m_i^2+m_j^2}{m_{\tilde{g}}
    }f_1(x_{d_k})-4 m_{\tilde{g}}f_2(x_{d_k})\right) \Biggr.\\ 
    &\Biggl. +\left(\Gamma_{DL}^{ki} \Gamma_{DL}^{kj*}
    \left(m_i^2+m_j^2\right)+2 \Gamma_{DR}^{ki} \Gamma_{DR}^{kj*} m_i 
    m_j\right)f_3(x_{d_k}) \Biggr\} \,,
  \end{split}
\end{equation}
\begin{equation}
 \begin{split}
 \delta Z_{ji}^{b,L,H}=&\frac{\alpha_s}{4 \pi} C_F \sum\limits_{k=1}^6 \left(\frac{1}{m_{\tilde{g}}} (\Gamma_{DL}^{kj*} \Gamma_{DR}^{ki} m_i + \Gamma_{DL}^{ki} \Gamma_{DR}^{kj*} m_j) f_1(x_{d_k}) + \Gamma_{DL}^{ki} \Gamma_{DL}^{kj*} f_3(x_{d_k})
 \right) \, ,
 \end{split}
\end{equation}
where the functions $f_i(x)$ read
\begin{equation}
 \begin{split}
  f_1(x)&= \frac{x^2-2 x \ln (x)-1}{(x-1)^3} \, , \\
  f_2(x)&= \frac{ x \ln (x) }{x-1}-1 \, , \\
  f_3(x)&=\frac{2x (x-2) \ln (x)-x^2+4x-3}{2 (x-1)^2 } \, .
 \end{split}
\end{equation}

\subsection{Diagonal case}
\label{subsection:quarkdiag}
For $j=i$ the on-shell renormalization conditions read
\begin{equation}\begin{split}\label{quarkdiagrencond}
& \left.\widetilde{\text{Re}}\,\Sigma_{ii}^{ren} \, u(p_i) \stackrel{!}{=} 0\right|_{p_i^2=m_i^2,} \\
&\left.\widetilde{\text{Re}}\,\bar{u}(p_i)\Sigma_{ii}^{ren}  \stackrel{!}{=} 0 \right|_{p_i^2=m_i^2,} \\
& \widetilde{\text{Re}}\,\frac{\pslash+m_i}{p^2-m_i^2} \, \Sigma_{ii}^{ren} \, u(p_i) \stackrel{!}{=} 0 \, , \\
&\widetilde{\text{Re}}\,\bar{u}(p_i) \, \Sigma_{ii}^{ren} \, \frac{\pslash +m_i}{p^2-m_i^2}
\stackrel{!}{=} 0 \,.
\end{split}\end{equation}
We note that these conditions do not uniquely fix the renormalization constants.
We are therefore free to choose the antihermitian parts $\delta Z_{ii}^{L,AH}$ and
$\delta Z_{ii}^{R,AH}$ to vanish:
\begin{equation}\begin{split} 
\delta Z_{ii}^{L,AH}=0 \, , \qquad \delta Z_{ii}^{R,AH}=0
\, .
\end{split}\end{equation}
This is the only choice where
the $\delta
m_i$ do not depend on the first derivative of 
$\Sigma_{ji}^L$, $\Sigma_{ji}^R$, $A_{ji}^L$, $A_{ji}^R$ with respect to
$p^2$; in other words, the $\delta m_i$ 
are only related to the pole position of the renormalized propagators,
as it should be. With this choice, the other renormalization constants
are uniquely fixed by the renormalization conditions (\ref{quarkdiagrencond}).

The results for
$\delta Z_{ji}^{L,H}$  ($\delta Z_{ji}^{R,H}$ is obtained by
exchanging $R\leftrightarrow L$) and for the mass shifts 
$\delta m_{i}$, again written in terms of 
$\Sigma_{ji}^L$, $\Sigma_{ji}^R$, $A_{ji}^L$ ,$A_{ji}^R$ and their
derivatives with respect to $p^2$ (denoted by a dot)  read
\begin{equation}\begin{split}
    & \delta m_{i} = \widetilde{\text{Re}}\,\left[\frac{m_i}{2}\left( \Sigma_{ii}^L(m_i^2) +
      \Sigma_{ii}^R(m_i^2) \right) + A_{ii}^L(m_i^2)\right],  \\
    & \delta Z_{ii}^{L,H} =\widetilde{\text{Re}}\, \Sigma_{ii}^L(m_i^2) +
    m_i \, \widetilde{\text{Re}}\,\left[ m_i \, \dot{\Sigma}_{ii}^L(m_i^2) +  m_i \,
      \dot{\Sigma}_{ii}^R(m_i^2) + \dot{A}_{ii}^L(m_i^2) +
      \dot{A}_{ii}^R(m_i^2) \right] .
\end{split}\end{equation}
These quantities are again needed up to
order $\epsilon^0$ and up to linear order in the quark mass
$m_i$. 

As already mentioned, the diagonal quantities 
$\Sigma_{ii}^L$, $\Sigma_{ii}^R$, $A_{ii}^L$  and $A_{ii}^R$ get
contributions both from squark-gluino loops and from quark-gluon
loops. We therefore write $\delta Z_{ii}^{L,H}$ and $\delta m_{i}$ as
a sum of two contributions: 
\begin{equation}\begin{split}
\delta Z_{ii}^{L,H} &= \delta {Z}_{ii}^{a,L,H} + \delta 
  Z_{ii}^{b,L,H} \, , \qquad  
\delta m_{i} = \delta m_{i}^{a} + \delta  m_{i}^{b} \, ,   
\end{split}\end{equation}
where the first and second term on the corresponding r.h.s.\ result from the
quark-gluon and the squark-gluino loops, respectively. 
Note that $\delta Z_{ii}^{a,L,H}$, $\delta Z_{ii}^{b,L,H}$,
$\delta m_{i}^{a}$ and $\delta m_{i}^{b}$  
contain $1/\epsilon$ poles. Explicitly, we get (using the Feynman
  gauge for the gluon propagator and $L_\mu=\ln (\mu^2/m_{\tilde{g}}^2)$)
\begin{equation}\begin{split}\label{quarkZ}
 \delta Z_{bb}^{a,L,H}=& \delta Z_{bb}^{a,R,H}=
- \frac{\alpha_s}{4 \pi} C_F \left(
\frac{1}{\epsilon} + \frac{2}{\epsilon_{{\rm ir}}} +4 + 3 \, \ln(\mu^2/m_b^2) \right), \\ 
\delta Z_{ss}^{a,L,H}=& \delta Z_{ss}^{a,R,H}= - \frac{\alpha_s}{4 \pi} C_F \left(\frac{1}{\epsilon}-\frac{1}{\epsilon_{{\rm ir}}}\right),\\
\delta Z_{ii}^{b,L,H}=&\frac{\alpha_s}{4 \pi} C_F \sum\limits_{k=1}^6 
\left(\frac{m_i}{m_{\tilde{g}}} (\Gamma_{DL}^{ki*} \Gamma_{DR}^{ki} + \Gamma_{DL}^{ki} \Gamma_{DR}^{ki*} ) f_1(x_{d_k}) + \Gamma_{DL}^{ki} \Gamma_{DL}^{ki*} f_3(x_{d_k})
 \right) \\
&- \frac{\alpha_s}{4 \pi} C_F \left( \frac{1}{\epsilon} +L_\mu \right)
\, .
\end{split}\end{equation}
\begin{equation}\begin{split}\label{quarkmi}
 \delta m_{i}^{a}= &
\frac{\alpha_s}{4 \pi} C_F \, m_i \, \left(\frac{3}{\epsilon}  +4 + 3 \, \ln(\mu^2/m_i^2) \right), \\ 
\delta m_{i}^{b}=  &
   \frac{1}{2} \, \frac{\alpha_s}{4 \pi} \, C_F \,
  \sum\limits_{k=1}^6 \left(
m_i (\Gamma_{DL}^{ki} \Gamma_{DL}^{ki*} + \Gamma_{DR}^{ki} \Gamma_{DR}^{ki*}) \, f_3(x_{d_k}) 
- 4 \, m_{\tilde{g}} \, \Gamma_{DL}^{ki} \Gamma_{DR}^{ki*} \, f_2(x_{d_k})  \right) \\
& - \frac{\alpha_s}{4 \pi} \, C_F \, m_i \,
\left(\frac{1}{\epsilon}+L_\mu \right) \, .
\end{split}\end{equation}
In (\ref{quarkZ}) the symbol $\epsilon_{\rm ir}$ has been
introduced in order to mark the infrared poles and distinguish them
from the ultraviolet ones.

\section{Squark-field and squark-mass renormalization in the on-shell scheme}
\label{sec:squarkrenormalization}
We start with the free (bare) Lagrangian for the down-type squarks
\begin{equation}\begin{split}
{\cal L} = - \tilde{\phi}_k^{0 \, \dagger} \delta_{kk'} \Box \tilde{\phi}_{k'}^{ 0} - 
(m_{\tilde{d}_k}^0)^2 \tilde{\phi}_k^{0 \, \dagger} \tilde{\phi}_k^0,
 \end{split}\end{equation}  
and express the bare quantities (marked with the superscript ``0'') in
terms of renormalized ones:
\begin{equation}\begin{split}
& \tilde{\phi}_k^0 = \left( \delta_{kk'} + \frac{1}{2} \delta  \tilde{
  Z}_{kk'} \right) \, \tilde{\phi}_{k'} \, ,  \\
& (m_{\tilde{d}_k}^0)^2 = (m_{\tilde{d}_k})^2 - \delta m_{\tilde{d}_k}^2 \, .
\end{split}\end{equation}
For the renormalized squark self-energies $\tilde{\Sigma}_{k'k}^{ren}$ we get
\begin{equation}\begin{split}
\label{squarkselfren}
\tilde{\Sigma}_{k'k}^{ren} =& \tilde{\Sigma}_{k'k} 
- \frac{1}{2} \left( \delta \tilde{Z} + \delta \tilde{
    Z}^\dagger \right)_{k'k} \, p^2 
+ \frac{1}{2}  (\delta \tilde{Z}^{\dagger})_{k'k}  m_{\tilde{d}_k}^2
+ \frac{1}{2} m_{\tilde{d}_{k'}}^2 (\delta \tilde{Z})_{k'k} -
\delta_{k'k} \, \delta m_{\tilde{d}_k}^2 \, ,
\end{split}\end{equation}
where $\tilde{\Sigma}_{k'k}$ denotes the bare squark self-energies.
 \begin{figure}
 \subfigure[]{
  \scalebox{0.7}{\begin{tabular}{c}\rule{0cm}{4.1cm}\\
	\includegraphics{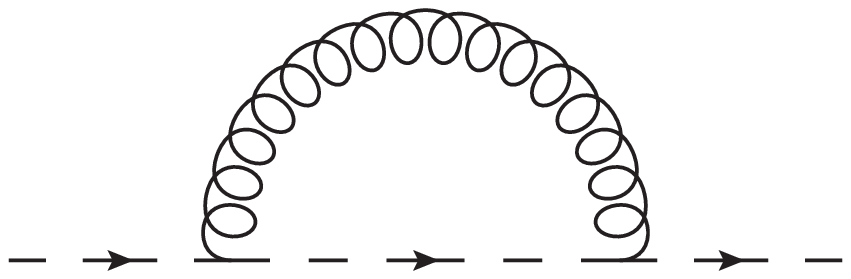}
\end{tabular}}
  }
   \subfigure[]{
  \scalebox{0.7}{\begin{tabular}{c}\includegraphics{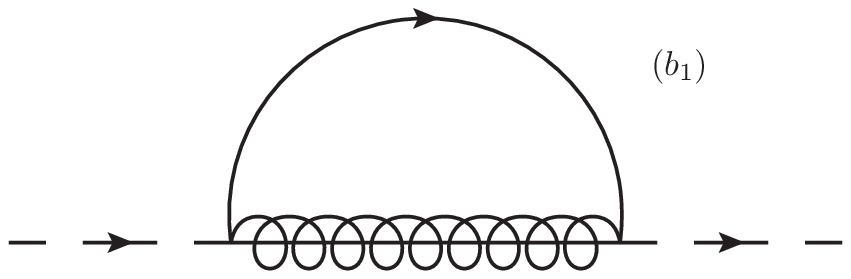}\\
  \includegraphics{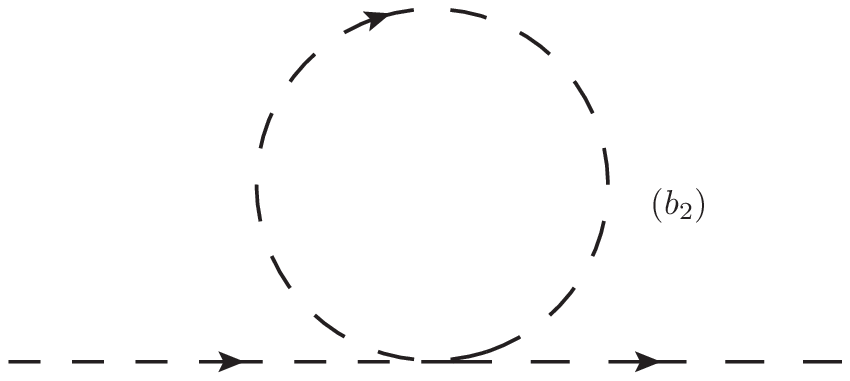}
  \end{tabular}}
  }\caption{Contributions to the bare squark self-energies $\tilde{\Sigma}_{k'k}$.}
\label{Fig:squarkself}
 \end{figure}
We note that for $k' \neq k$ the bare self-energies at order $\alpha_s^1$
receive contributions from diagram $(b_1)$ (see Figure \ref{Fig:squarkself}) with a quark and a gluino in the loop and
from diagram $(b_2)$ with a squark tadpole. In the
diagonal case $k'=k$ diagram (a) with an internal squark and a gluon
also contributes. 

\subsection{Off-diagonal case}
For $k' \neq k$ the on-shell renormalization conditions read
\begin{equation}\begin{split}
\label{squarkrencond}
\widetilde{\mbox{Re}}\, \Sigma_{k'k}^{ren}(p^2=m_{\tilde{d}_k}^2) \stackrel{!}{=} 0  \, , \qquad
\widetilde{\mbox{Re}}\, \Sigma_{k'k}^{ren}(p^2=m_{\tilde{d}_{k'}}^2) \stackrel{!}{=} 0  \, ,
\end{split}\end{equation}
where the symbol $\widetilde{\mbox{Re}}$ stands for taking  the dispersive part.
From these conditions and from (\ref{squarkselfren}) we get
\begin{equation}\begin{split}
\label{zsquarkoff}
\delta \tilde{Z}_{k'k} =& \frac{2}{m_{\tilde{d}_k}^2-m_{\tilde{d}_{k'}}^2}
\, \widetilde{\mbox{Re}}\, \tilde{\Sigma}_{k'k}(m_{\tilde{d}_k}^2) \, ,  \\
(\delta \tilde{Z}^\dagger)_{k'k} =& \frac{2}{m_{\tilde{d}_{k'}}^2-m_{\tilde{d}_k}^2}
\, \widetilde{\mbox{Re}}\, \tilde{\Sigma}_{k'k}(m_{\tilde{d}_{k'}}^2) \, .
\end{split}\end{equation}
When looking at the explicit result for the off-diagonal bare
self-energies 
$\tilde{\Sigma}_{k'k}$ below,
one finds that these two equations are compatible with each other.
 Therefore only one of them is needed.

The off-diagonal bare self-energies $\tilde{\Sigma}_{k'k}(p^2)$ can
be decomposed as
\begin{equation}\begin{split}
\label{eq:3}
\tilde{\Sigma}_{k'k}(p^2) = \tilde{\Sigma}_{k'k}^{(b_1)}(p^2) 
+ \tilde{\Sigma}_{k'k}^{(b_2)} \, ,
\end{split}\end{equation}
where $\tilde{\Sigma}_{k'k}^{(b_1)}(p^2)$ and  $\tilde{\Sigma}_{k'k}^{(b_2)}$
correspond to diagrams $(b_1)$ and $(b_2)$ in Figure \ref{Fig:squarkself}. 
As indicated in the notation, $\tilde{\Sigma}_{k'k}^{(b_2)}$ is independent
of the external momentum.
Explicitly, one obtains for
$\tilde{\Sigma}_{k'k}^{(b_1)}(p^2)$ (retaining only linear terms in the down-type
quark mass $m_d$)
\begin{equation}\begin{split}
\tilde{\Sigma}_{k'k}^{(b_1)}(p^2) =& - \frac{\alpha_s}{\pi} C_F \,
\left[ \sum_d \left(
m_{\tilde{g}} \, m_d \, (\Gamma_{DR}^{kd*} \Gamma_{DL}^{k'd}+ \Gamma_{DL}^{kd*}
\Gamma_{DR}^{k'd}) \right) \right]   \\
& \times
\left[ \frac{1}{\epsilon} + 2 + L_\mu +
\left(\frac{m_{\tilde{g}}^2}{p^2} - 1 \right) \,  \ln \left( 1- \frac{p^2}{m_{\tilde{g}}^2} - i \delta \right)
\right]   \\
&- \frac{\alpha_s}{2\pi} C_F \, \delta_{kk'} \, \left[ \frac{1}{\epsilon} 
(p^2-2 m_{\tilde{g}}^2) + 2 p^2 -3 m_{\tilde{g}}^2 + (p^2 - 2
  m_{\tilde{g}}^2) L_\mu 
\right.  \\
&
\left.
+ \left( 2 m_{\tilde{g}}^2 - p^2 - \frac{m_{\tilde{g}}^4}{p^2} \right) \, 
\ln \left( 1- \frac{p^2}{m_{\tilde{g}}^2} - i \delta \right) \right] \, ,
\label{eq::sigtil}
\end{split}\end{equation}
while 
$\tilde{\Sigma}_{k'k}^{(b_2)}$ reads (using $X_D^{kk'} =
\Gamma_{DR}^{ki} \Gamma_{DR}^{k'i*} - 
\Gamma_{DL}^{ki} \Gamma_{DL}^{k'i*}$ with $i$ summed over $i=1,2,3$)
\begin{equation}
\tilde{\Sigma}_{k'k}^{(b_2)} = - \frac{ \alpha_s}{3\pi} \, 
X_D^{k'l} \, m_{\tilde{d}_l}^2 \, X_D^{lk} \, \left[ \frac{1}{\epsilon} +1 + \ln \frac{\mu^2}{m_{\tilde{d}_l}^2}
\right] \, .
\end{equation}
These expressions for $\tilde{\Sigma}_{k'k}^{(b_1)}$ and
$\tilde{\Sigma}_{k'k}^{(b_2)}$ hold for both  the off-diagonal and the
diagonal case.
The explicit result for $\delta \tilde{Z}_{k'k}$ reads (for $k'
\neq k$, the label $b$ indicates that there are only contributions from class b))
\begin{equation}\begin{split}
 \delta \tilde{Z}^b_{k'k} =& - \frac{\alpha_s}{\pi} \,
\frac{2}{x_{d_k}-x_{d_{k'}}} \, \left[
 C_F \, \left[ \sum_d \left(
\frac{m_d}{m_{\tilde{g}}} \, (\Gamma_{DR}^{kd*} \Gamma_{DL}^{k'd}+ \Gamma_{DL}^{kd*}
\Gamma_{DR}^{k'd}) \right) \right]   \right. \\
& \times
\left( \frac{1}{\epsilon} + 2 + L_\mu -
\frac{x_{d_k}-1}{x_{d_k}} \,  \ln \left| 1- x_{d_k} \right|
\right)  \\
& \left. + \frac{1}{3}
X_D^{k'l} \, x_{d_l} \, X_D^{lk} \, \left( \frac{1}{\epsilon} +1 + \ln \frac{\mu^2}{m_{\tilde{d}_l}^2}
\right) \right] \, .
\end{split}\end{equation}
The hermitian and antihermitian parts are:
\begin{equation}\begin{split}
 \delta \tilde{Z}_{k'k}^{b,H} =& - \frac{\alpha_s}{\pi} \, C_F \, 
\frac{2}{x_{d_k}-x_{d_{k'}}} \, 
  \left[ \sum_d \left(
\frac{m_d}{m_{\tilde{g}}} \, (\Gamma_{DR}^{kd*} \Gamma_{DL}^{k'd}+ \Gamma_{DL}^{kd*}
\Gamma_{DR}^{k'd}) \right) \right]    \\
&  \times
\left(   \frac{x_{d_{k'}}-1}{2x_{d_{k'}}} \,  \ln \left| 1- x_{d_{k'}} \right|
        - \frac{x_{d_k}-1}{2x_{d_k}} \,  \ln \left| 1- x_{d_k} \right|
\right)  \, ,
\end{split}\end{equation}
\begin{equation}\begin{split}
 \delta \tilde{Z}_{k'k}^{b,AH} =& - \frac{\alpha_s}{\pi} \,
\frac{2}{x_{d_k}-x_{d_{k'}}} \, \left\{
 C_F \, \left[ \sum_d \left(
\frac{m_d}{m_{\tilde{g}}} \, (\Gamma_{DR}^{kd*} \Gamma_{DL}^{k'd}+ \Gamma_{DL}^{kd*}
\Gamma_{DR}^{k'd}) \right) \right]   \right. \\
& \times
\left( \frac{1}{\epsilon} + 2 + L_\mu 
- \frac{x_{d_{k'}}-1}{2x_{d_{k'}}} \,  \ln \left| 1- x_{d_{k'}} \right|
- \frac{x_{d_k}-1}{2x_{d_k}} \,  \ln \left| 1- x_{d_k} \right|
\right)  \\
& \left. + \frac{1}{3}
X_D^{k'l} \, x_{d_l} \, X_D^{lk} \, \left( \frac{1}{\epsilon} +1 + \ln \frac{\mu^2}{m_{\tilde{d}_l}^2}
\right) \right\} \, ,
\end{split}\end{equation}
where we used the shorthand notation
$x_{d_k}=m_{\tilde{d}_k}^2/m_{\tilde{g}}^2$. The result for $\delta
\tilde{Z}_{kk}$ is given in the following subsection.

\subsection{Diagonal case}\label{subsection:squarkdiag}
For $k'=k$ the on-shell renormalization conditions read
\begin{equation}\begin{split}\label{squarkdiagrencond}
 \widetilde{\mbox{Re}}\,\Sigma_{kk}^{ren}(p^2)  &\stackrel{!}{=} 0 , 
\mbox{ for $p^2 \to m_{\tilde{d}_k}^2$} , \\
 \frac{1}{p^2-m_{\tilde{d}_k}^2} \, \widetilde{\mbox{Re}}\,\Sigma_{kk}^{ren}(p^2) &\stackrel{!}{=} 0 ,
\mbox{ for $p^2 \to m_{\tilde{d}_k}^2$} .
\end{split}\end{equation}
These conditions fix  $\delta m_{\tilde{d}_k}^2$ to be
\begin{equation}\begin{split}
\delta m_{\tilde{d}_k}^2 =
\widetilde{\mbox{Re}}\,\tilde{\Sigma}_{kk}(m_{\tilde{d}_k}^2) \, . 
\end{split}\end{equation}  
However, they only fix the hermitian part (real part)
of $\delta \tilde{Z}_{kk}$. We choose the antihermitian part
(imaginary part) to vanish and get
\begin{equation}\begin{split}
\delta \tilde{Z}_{kk} = \left. \frac{\partial
  \widetilde{\mbox{Re}}\,\tilde{\Sigma}_{kk}(p^2) }{\partial p^2}
\right|_{p^2 \to m_{\tilde{d}_k}^2} \, .
\end{split}\end{equation}
As already mentioned, the diagonal bare squark self-energies
$\tilde{\Sigma}_{kk}$ get contributions from diagrams~(a) and
diagram (b) in Figure \ref{Fig:squarkself}. It turns out to be
convenient to decompose 
$\delta m^2_{\tilde{d}_k}$ and
$\delta \tilde{Z}_{kk}$ accordingly:
\begin{equation}\begin{split}
\delta \tilde{Z}_{kk} =
\delta \tilde{Z}_{kk}^{a} +
\delta \tilde{Z}_{kk}^{b} \, , \qquad
\delta m_{\tilde{d}_k}^2 =
\delta m_{\tilde{d}_k}^{2,a} +
\delta m_{\tilde{d}_k}^{2,b} \, . 
\end{split}\end{equation}
Explicitly, we find
\begin{equation}\begin{split}
\delta m_{\tilde{d}_k}^{2,a} =&  \frac{\alpha_s}{4 \pi} C_F \,
m_{\tilde{d}_k}^2 \,  \left(
\frac{3}{\epsilon} + 7 + 3 \, \ln \frac{\mu^2}{m_{\tilde{d}_k}^2} \right),
\end{split}\end{equation}
\begin{equation}\begin{split}
\delta m_{\tilde{d}_k}^{2,b} =& - \frac{\alpha_s}{\pi} C_F \, \left[
  \sum_d \left(
m_{\tilde{g}} \, m_d \, (\Gamma_{DR}^{kd*} \Gamma_{DL}^{kd}+ \Gamma_{DL}^{kd*}
\Gamma_{DR}^{kd}) \right) \right]
\left[ \frac{1}{\epsilon} + 2 + L_\mu -
\frac{x_{d_k}-1}{x_{d_k}} \,  \ln \left| 1- x_{d_k} \right|
\right]   \\
&- \frac{\alpha_s}{2\pi} C_F \, m_{\tilde{g}}^2  \left[ \frac{1}{\epsilon} 
(x_{d_k}-2) + 2 x_{d_k} -3  + (x_{d_k} - 2) L_\mu -
\frac{(x_{d_k}^2 - 2 x_{d_k} +1)}{x_{d_k}} \, 
\ln \left| 1- x_{d_k} \right| \right]  \\
&
- \frac{\alpha_s}{3\pi} \, m_{\tilde{g}}^2 \,
X_D^{kl} \, x_{d_l} \, X_D^{lk} \, \left[ \frac{1}{\epsilon} +1 + \ln \frac{\mu^2}{m_{\tilde{d}_l}^2} \right] \, ,
\end{split}\end{equation}

\begin{equation}\begin{split}\label{squarkZb}
\delta \tilde{Z}_{kk}^{a} =&  \frac{\alpha_s}{4 \, \pi} \, C_F \,
\left[ \frac{2}{\epsilon} - \frac{2}{\epsilon_{{\rm ir}}} \right] \, ,
\end{split}\end{equation}
\begin{equation}\begin{split}\label{squarkZa}
\delta \tilde{Z}_{kk}^{b} =& - \frac{\alpha_s}{2 \, \pi} C_F \left[ \frac{1}{\epsilon} +
2+ L_\mu - \frac{(x_{d_k}^2-1)}{x_{d_k}^2} \ln \left| 1-x_{d_k} \right| - 
\frac{(2 x_{d_k}-x_{d_k}^2-1)}{x_{d_k} \, (1-x_{d_k})} \right]   \\
&
+ \frac{\alpha_s}{\pi} C_F \, \left[ \sum_d \left(
\frac{m_d}{ m_{\tilde{g}}} \, (\Gamma_{DR}^{*kd} \Gamma_{DL}^{kd}+ \Gamma_{DL}^{*kd}
\Gamma_{DR}^{kd}) \right) \right] \, \frac{\left( x_{d_k} +  
\ln \left| 1 - x_{d_k} \right| \right)}{x_{d_k}^2} \, .
\end{split}\end{equation}

\section{Gluino-field and Gluino-mass renormalization in the on-shell scheme}\label{sec:gluinorenormalization}
We start with the free (bare) Lagrangian for a gluino, which is
described by a four-component Majorana field $\psi_{\tilde{g}}^0$:
\begin{equation}\begin{split}
{\cal L} = \frac{i}{2} \, \bar{\psi}_{\tilde{g},L}^0 \, \gamma^\mu \partial_\mu 
\psi_{\tilde{g},L}^0 
         + \frac{i}{2} \, \bar{\psi}_{\tilde{g},R}^0 \, \gamma^\mu \partial_\mu 
 \psi_{\tilde{g},R}^0 - \frac{1}{2} \, \bar{\psi}_{\tilde{g},R}^0 m_{\tilde{g}}^0 \psi_{\tilde{g},L}^0
- \frac{1}{2} \, \bar{\psi}_{\tilde{g},L}^0 (m_{\tilde{g}}^0)^* \psi_{\tilde{g},R}^0 \, ,
 \end{split}\end{equation}  
and express the bare quantities (marked with the superscript $(0)$) in
terms of renormalized ones:
\begin{equation}\begin{split}
& \psi_{\tilde{g},L}^0 = \left( 1 + \frac{1}{2} \delta  \tilde{  Z}_3^{L} \right) \, \psi_{\tilde{g},L} \, , \qquad
\psi_{\tilde{g},R}^0 = \left( 1 + \frac{1}{2} \delta  \tilde{  Z}_3^{R} \right) \, \psi_{\tilde{g},R} \, ,  \\
& m_{\tilde{g}}^0 = m_{\tilde{g}} - \delta m_{\tilde{g}} \, .
\end{split}\end{equation}
Due to the Majorana nature of the gluino, we have 
$\psi_{\tilde{g},R}^{(0)} = \left(\psi_{\tilde{g},L}^{(0)}\right)^c$ and
therefore $\delta \tilde{Z}_3^R = (\delta \tilde{Z}_3^L)^*$.
For the renormalized gluino self-energy $\Sigma^{ren}$, we get
 \begin{figure}
 \subfigure[]{
  \scalebox{0.7}{\begin{tabular}{c}\rule{0cm}{1.8cm}\\
        \includegraphics{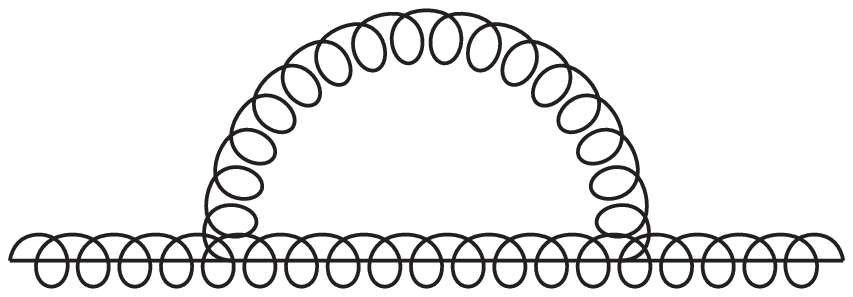}
\end{tabular}}
  }
  \subfigure[]{
  \scalebox{0.7}{\begin{tabular}{c}\includegraphics{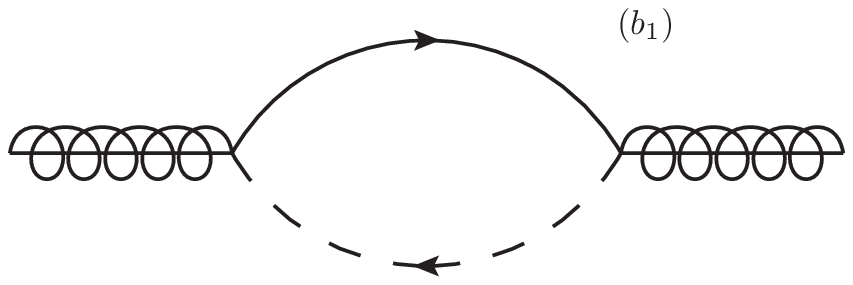}\\
  \includegraphics{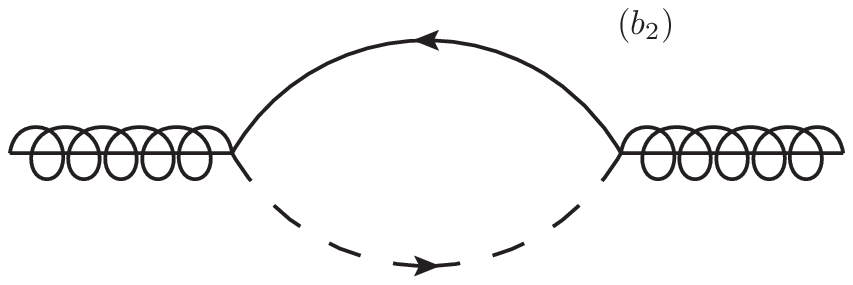}
  \end{tabular}}
  }\caption{Contributions to the bare gluino self-energy $\Sigma$.}
\label{Fig:gluinoself}
 \end{figure}
\begin{equation}\begin{split}
\Sigma^{ren} =& \Sigma - \frac{1}{2} \, (\delta \tilde{Z}_3^L +
\delta \tilde{Z}_3^{L,*}) \, \pslash 
+ m_{\tilde{g}} \, \delta \tilde{Z}_3^L \, P_L 
+ m_{\tilde{g}} \, \delta \tilde{Z}_3^{L,*} \, P_R 
- \delta m_{\tilde{g}} \, P_L
- \delta m_{\tilde{g}}^* \, P_R \, .
\end{split}\end{equation}
The bare gluino self-energy $\Sigma$, which gets contributions
from the three diagrams in Figure~\ref{Fig:gluinoself}, can be decomposed as
\begin{equation}
\Sigma = \pslash \, P_L \, \Sigma^L(p^2) + \pslash \, P_R \,
\Sigma^R(p^2) + P_L \, A^L(p^2)+ P_R \, A^R(p^2) \, .
\end{equation}
The on-shell renormalization conditions are formulated in the same way as for
the quarks in the diagonal case $j=i$ in Subsection~\ref{subsection:quarkdiag}.
The generic formulae for the hermitian and antihermitian part of
$\tilde{Z}_3^L$ and for the mass shift $\delta m_{\tilde{g}}$ read
\begin{equation}\begin{split}
& \delta \tilde{Z}_3^{L,H} = \Sigma^L(m_{\tilde{g}}^2) + 2 \,
m_{\tilde{g}}^2 \, \dot{\Sigma}^L(m_{\tilde{g}}^2) +
m_{\tilde{g}} \, \dot{A}^L(m_{\tilde{g}}^2) +
m_{\tilde{g}} \, \dot{A}^R(m_{\tilde{g}}^2) \, ,  \\
& \delta \tilde{Z}_3^{L,AH} = \frac{1}{2\, m_{\tilde{g}}} \,
\left(A^R(m_{\tilde{g}}^2) - A^L(m_{\tilde{g}}^2) \right) \, ,  \\
& \delta m_{\tilde{g}} = m_{\tilde{g}} \, \Sigma^L(m_{\tilde{g}}^2)+
\frac{1}{2} \left[ A^L(m_{\tilde{g}}^2) + A^R(m_{\tilde{g}}^2) \right] \, .
\end{split}\end{equation}
We now turn to the explicit results which we decompose 
as
\begin{equation}\begin{split}
\delta m_{\tilde{g}} = \delta m_{\tilde{g}}^a + \delta m_{\tilde{g}}^b
\, ,\qquad
\delta \tilde{Z}_3^{L} =
\delta \tilde{Z}_3^{a,L} +
\delta \tilde{Z}_3^{b,L} \, ,
\end{split}\end{equation}
according to the contributions shown in Figure \ref{Fig:gluinoself}.
We find (using the Feynman gauge for the gluon propagator)
\begin{equation}\begin{split}\label{gluinoZ}
\delta \tilde{Z}^{a,L,H}_3 =& -\frac{\alpha_s}{4\pi} \, C_A \, 
\frac{1}{\epsilon} +\mbox{finite},  \\
\delta \tilde{Z}^{a,L,AH}_3 =&\, 0, \\
\delta \tilde{Z}^{b,L,H}_3 =& -\frac{\alpha_s}{4\pi} \, 2 \, tr \, 
\frac{n_f}{\epsilon} +\mbox{finite} , \\
\delta \tilde{Z}^{b,L,AH}_3 =& \,\mbox{finite} \, ,  
\end{split}\end{equation}
where `finite' denotes ultraviolet finite pieces not needed
in this work. Numerically, we have $tr=1/2$, $C_A=3$, $n_f=6$.
For the mass shift $\delta m_{\tilde{g}}$ the explicit
results read in the on-shell scheme 
\begin{equation}\begin{split}
\label{deltamg}
\delta m_{\tilde{g}}^{a} =& \frac{\alpha_s}{4\pi} \, C_A \,
m_{\tilde{g}} \, \left[
\frac{3}{\epsilon} + 3 \, \ln \frac{\mu^2}{m_{\tilde{g}}^2} + 4 \right] \, ,\\
\delta m_{\tilde{g}}^{b} =& -\frac{\alpha_s}{4\pi} \, 2 \, tr \,
\frac{n_f}{\epsilon} m_{\tilde{g}} 
+  \frac{\alpha_s}{4\pi} \,  tr \, m_{\tilde{g}} \sum_{k,i,Q}
\left(\Gamma_{QL}^{ki} \, \Gamma_{QL}^{ki*} + \Gamma_{QR}^{ki} \,
\Gamma_{QR}^{ki*} \right)   \left[ 
  \frac{m_{Q_i}^2}{m_{\tilde{g}}^2} \left(1+\ln \frac{\mu^2}{m_{Q_i}^2} \right) \right. \\
& \left.-
\frac{m_{\tilde{Q}_k}^2}{m_{\tilde{g}}^2} \left(1+ \ln
\frac{\mu^2}{m_{\tilde{Q}_k}^2} \right) +
\frac{m_{\tilde{Q}_k}^2 - m_{Q_i}^2 - m_{\tilde{g}}^2}{m_{\tilde{g}}^2} \,
\widetilde{\mbox{Re}}\hat{B}_0(m_{\tilde{g}}^2;m_{\tilde{Q}_k},m_{Q_i}) \right]  \\
& + \frac{\alpha_s}{4\pi} \, 2 \, tr \, \sum_{k,i,Q} m_{Q_i} \,
\left(\Gamma_{QL}^{ki} \, \Gamma_{QR}^{ki*} + \Gamma_{QR}^{ki} \,
\Gamma_{QL}^{ki*} \right) \,
\widetilde{\mbox{Re}}\hat{B}_0(m_{\tilde{g}}^2;m_{\tilde{Q}_k},m_{Q_i}) \, .
\end{split}\end{equation}
The indices in the sums within the expression for  $\delta
m_{\tilde{g}}^{b}$ run over the following ranges: $k=1,...,6$,  $i=1,2,3$, and $Q=D,U$.
For $Q=D$ ($Q=U$) the symbols $m_{Q_i}$ and $m_{\tilde{Q}_k}$ denote the
masses of the down-type (up-type) quarks and down-type (up-type)
squarks, respectively.
The function $\hat{B}_0$ appearing in  (\ref{deltamg}) is defined as
\begin{equation}\begin{split}
\hat{B}_0(p^2;m_1,m_2) = -\int_0^1 \, dx \, \ln \frac{-p^2 \, x \, (1-x) +
  m_1^2 \, x +m_2^2 \, (1-x) - i \delta }{\mu^2} \, . 
\end{split}\end{equation}

\section{Gluon-field renormalization} \label{sec:gluonrenormalization} 
\begin{figure}
\begin{center}
 \subfigure[]{
  \scalebox{0.5}{\begin{tabular}{c}\includegraphics{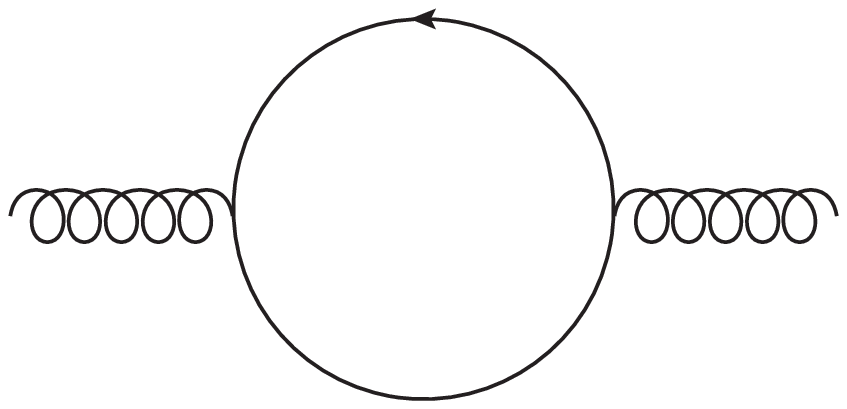}\\
        \includegraphics{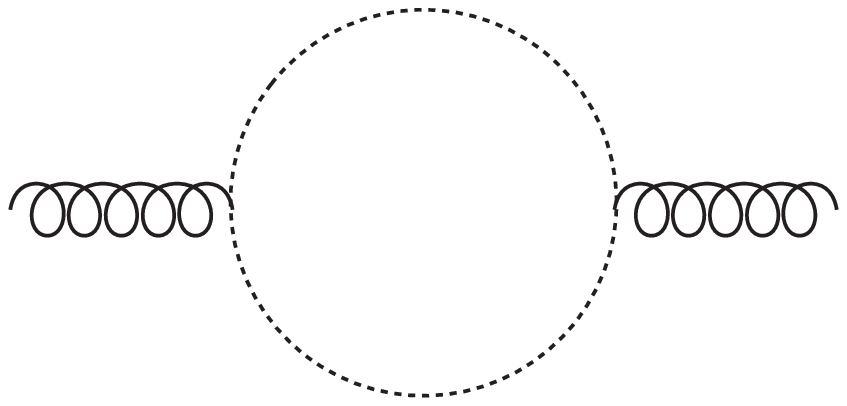} \\
        \includegraphics{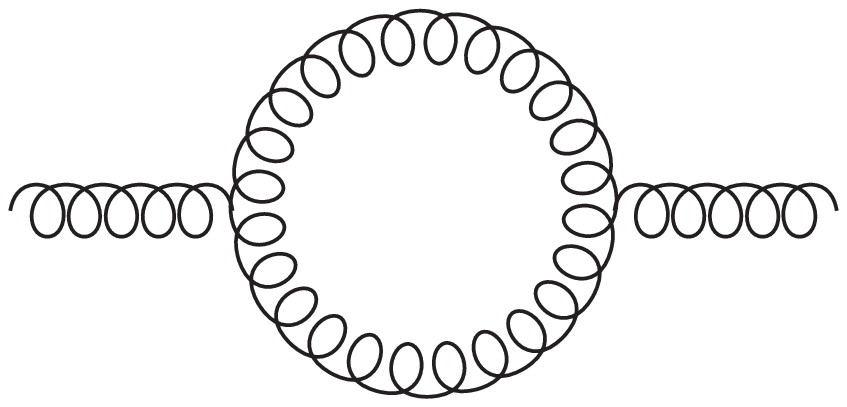}
\end{tabular}}
  }
   \subfigure[]{
  \scalebox{0.5}{\begin{tabular}{c}\rule{0cm}{4.3cm}\\
  \includegraphics{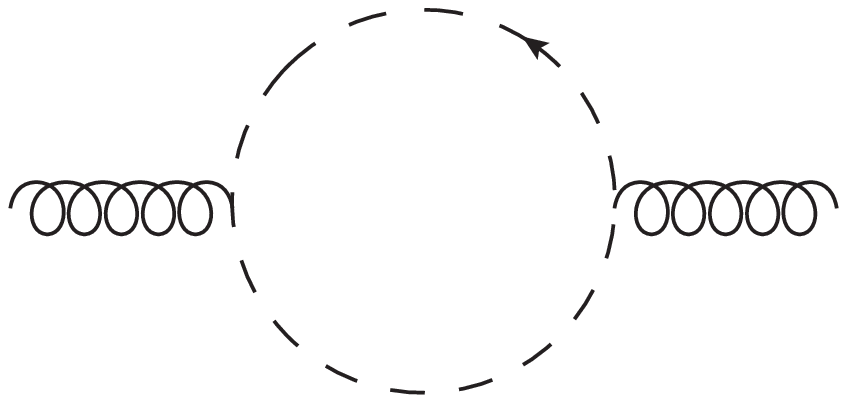}\\
  \includegraphics{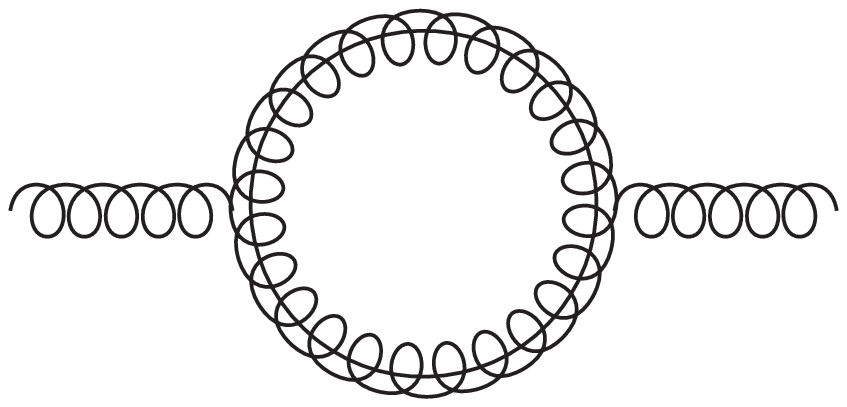}
  \end{tabular}}
  }\caption{Contributions to the bare gluino two-point function.}\label{Fig:gluonself}
  \end{center}
 \end{figure}
To fix the Wilson coefficients of the chromomagnetic operators 
${\cal O}_{8}$, which are related to the process $b \to s g$, 
we need the on-shell renormalization constant
(LSZ-factor) for the gluon field.
The diagrams which define the bare gluon two-point function at order
$\alpha_s$ are shown in Figure~\ref{Fig:gluonself}. The corresponding
gluon field renormalization constant $Z_3$, defined 
through $A_\mu^0=\sqrt{Z_3} \, \, A_\mu$ (where $A_\mu^0$ and
$A_\mu$ denote the bare and renormalized gluon field, respectively),
can be written as
\begin{equation}\begin{split}
Z_3 = 1 + \delta Z_3^a + \delta Z_3^b \, ,
\end{split}\end{equation}
according to the contributions shown in Figure \ref{Fig:gluonself}.
As we do not want to reproduce the order $\alpha_s$ correction to the Wilson
coefficients $C_8$ due to {\it gluons} (which were computed in
\cite{Bobeth:1999ww}), we only need $\delta Z_3^b$ in our work. We obtain
\begin{equation}\begin{split}
\label{deltaZ3aonshell}
 \delta Z_3^b = \frac{\alpha_s}{4\pi} \frac{2}{3} \left[
   \frac{1}{2} \, tr \, \left( 
\sum \limits_{k=1}^{6} \ln \frac{m^2_{\tilde{d}_k}}{m^2_{\tilde{g}}} +
\sum \limits_{k=1}^{6} \ln \frac{m^2_{\tilde{u}_k}}{m^2_{\tilde{g}}}
 - 2 \, n_f \ln
\frac{\mu^2}{m_{\tilde{g}}^2}-\frac{2 \, n_f}{\epsilon} \right) - C_A
\left( \frac{1}{\epsilon} + \ln
\frac{\mu^2}{m_{\tilde{g}}^2}\right)\right] \, ,
\end{split}\end{equation}
where $n_f=6$.

\section{Renormalization of $g_{s,Y}$, $\Gamma_{DL/R}^{ki}$ 
and $g_{s,G}$ in the full theory} \label{renormg}
\subsection{Renormalization of $g_{s,Y}$ and $\Gamma_{DL/R}^{ki}$ }\label{gYren}
We use the squark-quark-gluino-vertex 
in order to compute the one-loop counterterms for the
strong coupling constant  $g_{s,Y}$ of Yukawa type 
and the $\Gamma_{DL/R}$ matrices
in the $\overline{\rm MS}$ scheme.
At tree-level, this vertex for a down-type squark of flavor $k$ (color
$\beta$), a down-type quark of flavor $i$ (color $\alpha$) and a
gluino (color $a$) reads (see \eqref{sqg_vertex1})
\begin{equation}
 V^{tree}_{ki}=-ig_s\sqrt{2}T^a_{\beta\alpha}
\left(\Gamma_{DL}^{ki} P_L - \Gamma_{DR}^{ki} P_R \right) \, .
\end{equation}
At one-loop order, there are four one-particle irreducible diagrams 
contributing to the vertex correction. Three of them contain a
virtual gluon in the loop,
while the remaining one contains a quark-squark-gluino loop.
It turns out that the latter contribution is finite (after using  the 
unitarity relations) and thus there is no contribution
in the $\overline{\rm MS}$ scheme. For the
vertex correction $\delta V^{irred}_{ki}$ due to these irreducible diagrams
we obtain (retaining only the UV-singularities)
\begin{equation}
  \delta V^{irred}_{ki} = \frac{\alpha_s}{4\pi} \, \frac{2
    C_A+C_F}{\epsilon}  \,
  V^{tree}_{ki} \, .
\end{equation}
Additionally, there is a counterterm contribution $\delta V^{ct}_{ki}$
which contains the following sources: renormalization of the quark,
squark and gluino fields, as well as the renormalization of the parameters 
$g_{s,Y}$ and $\Gamma_{DL/R}^{ki}$. In the
$\overline{\rm MS}$ scheme the renormalization constants for the
  various fields 
can be extracted from the previous sections in this appendix. Splitting
them again into class~a) and class~b) contributions, the non-vanishing
ones read for the quarks
\begin{equation}
\begin{split}
 \delta Z^a &\equiv \delta Z^{a,L}_{ii} = \delta Z^{a,R}_{ii} = - \frac{\alpha_s}{4\pi} \,
C_F \, \frac{1}{\epsilon} \, ,  \\
 \delta Z^b& \equiv \delta Z^{b,L}_{ii} = \delta Z^{b,R}_{ii} = - \frac{\alpha_s}{4\pi} \, C_F \, \frac{1}{\epsilon}\,,
\end{split}
\end{equation}
and for the squarks
\begin{equation}
\label{squarkrenminimal}
\begin{split}
 \delta \tilde{Z}^a \equiv \,& \delta \tilde{Z}^{a}_{kk} =  \frac{\alpha_s}{4\pi} \, C_F \,
\frac{2}{\epsilon} \, ,\\
\delta \tilde{Z}^b \equiv \,&\delta \tilde{Z}^{b}_{kk} = - \frac{\alpha_s}{4\pi} \, C_F \,
\frac{2}{\epsilon} \, ,  \\
\delta \tilde{Z}^{b,AH}_{kk'} = &
 - \frac{\alpha_s}{\pi} \,
\frac{2}{x_{d_{k'}}-x_{d_{k}}} \, \left[
 C_F \,  \sum_d \left(
\frac{m_d}{m_{\tilde{g}}} \, ( \Gamma_{DL}^{kd} \Gamma_{DR}^{k'd*} + 
 \Gamma_{DR}^{kd} \Gamma_{DL}^{k'd*} ) \right) \right.\\
 &\left.+ \frac{1}{3}
X_D^{kl} \, x_{d_l} \, X_D^{lk'} \right] \,  \frac{1}{\epsilon}, \qquad (k \ne k'),
\end{split}
\end{equation}
and finally for the gluino 
($\tilde{Z}_3 = 1 + \delta \tilde{Z}_3^a + \delta \tilde{Z}_3^b$)
\begin{equation}
 \delta \tilde{Z}^{a}_{3} = - \frac{\alpha_s}{4\pi} \, C_A \,
\frac{1}{\epsilon} \, , \qquad
\delta \tilde{Z}^{b}_{3} = - \frac{\alpha_s}{4\pi} \, 2 \, tr \, n_f \, \frac{1}{\epsilon}.
\end{equation}
The counterterm contribution to the vertex due to the renormalization
of the fields can be written as
\begin{equation}
\begin{split}
 \delta V^{ct,fields}_{ki} = &
\frac{1}{2} \left( \delta Z^a + \delta Z^b + \delta \tilde{Z}^a +
\delta \tilde{Z}^b + \delta \tilde{Z}^a_3 + \delta \tilde{Z}^b_3
\right) V_{ki}^{tree}\\
&- \frac{1}{2} \sum_{k'} \, \delta \tilde{Z}_{kk'}^{b,AH} \, V_{k'i}^{tree} \, . 
\end{split}
\end{equation}
We now turn to the renormalization of the parameters $g_{s,Y}$
and $\Gamma_{DL/R}$ matrices. For the former we write the connection
between the bare- and renormalized version in the form
$g_{s,Y}^{bare}=(1+\delta Z_{g_s,Y}^a +\delta Z_{g_s,Y}^b) \, g_{s,Y}$
while for the latter the corresponding connection reads 
\begin{equation}
\Gamma_{DL/R}^{bare,ki} = \sum_{k'} \left( \delta^{kk'} + \delta \Gamma_D^{kk'} \right)
\Gamma_{DL/R}^{k'i} \, ,
\end{equation}
where $\delta \Gamma_D^{kk'}$ is an antihermitian $6 \times 6$ matrix.
This procedure corresponds to a rotation of the squarks
only. Therefore the renormalized $\Gamma_{DL/R}$ matrices still describe the transition between the super-CKM basis and the mass
eigenstate basis.  
The counterterm contribution to the vertex due to the renormalization
of the parameters then reads
\begin{equation}
\begin{split}
& \delta V^{ct,param}_{ki} = 
 \delta Z_{g_s,Y} V_{ki}^{tree} + 
 \sum_{k'} \, \delta \Gamma_{D}^{kk'} \, V_{k'i}^{tree} \, . 
\end{split}
\end{equation}
Requiring $\delta V_{ki}^{irred}+\delta V_{ki}^{ct,fields}+\delta V_{ki}^{ct,param}$
to be finite, fixes 
$\delta Z_{g_s,Y}^a$,
$\delta Z_{g_s,Y}^b$ and
$\delta \Gamma_{D}^{kk'}$ 
in the $\overline{\rm MS}$ scheme to be
($Z_{g_s,Y} = 1 + \delta Z_{g_s,Y}^a + \delta Z_{g_s,Y}^b $)
\begin{equation}
 \begin{split}
  \delta Z_{g_s,Y}^{a} &=-\frac{\alpha_s}{4\pi} \frac{1}{\epsilon}\frac{3}{2}(C_A+ C_F) ,\\
  \delta Z_{g_s,Y}^{b} &= \frac{\alpha_s}{4\pi}
  \frac{1}{\epsilon}\left( \frac{3}{2} C_F+ tr\, n_f \right),\\
  \delta \Gamma_D^{kk'} &= \frac{1}{2} \delta \tilde{Z}_{kk'}^{b,AH}  \, ,
 \end{split}
\end{equation}
where  $\delta \tilde{Z}_{kk'}^{b,AH}$ is given in (\ref{squarkrenminimal}).

\subsection{Renormalization of $g_{s,G}$}\label{gGren}
We use the squark-squark-gluon-vertex 
to compute the one-loop renormalization constant for the
strong coupling constant  $g_{s,G}$ of gauge type 
in the $\overline{\rm MS}$ scheme.

At the one-loop order, there are
six one-particle irreducible diagrams contributing to the vertex
correction.
Two of them contain a gluino and a quark in the loop, one consists of a squark loop, while the
remaining three feature a virtual gluon. For the vertex correction
$\delta \tilde{V}_{kk}^{irred}$ due to these irreducible diagrams we
obtain (keeping only the UV-singularities)
\begin{equation}\begin{split}
\delta \tilde{V}_{kk}^{irred} =& - \frac{\alpha_s}{4\pi} \, (2 \, C_F
- C_A) \, \frac{1}{\epsilon} \, V_{kk}^{tree} 
 + \frac{\alpha_s}{4\pi} \, 2 \, C_F \, \frac{1}{\epsilon} \, V_{kk}^{tree}  
  \, ,
\end{split}\end{equation}
where the first term on the r.h.s. is due to gluon
corrections, while the second one results from gluinos and squarks. 
Additionally, there is a counterterm contribution $\delta \tilde{V}^{ct}_{kk}$
which contains the following sources: renormalization of the 
squark and the gluino field, as well as the renormalization of 
$g_{s,G}$.  
Again, the renormalization constants for the
various fields in the $\overline{\rm MS}$ scheme 
can be extracted from the previous sections in this appendix. 
Unlike for the squark-quark-gluino-vertex, only the diagonal
renormalization constants for the squark fields are involved in the squark-squark-gluon-vertex.  
The gluon-field renormalization constant reads
($Z_3= 1 + \delta Z_3^a + \delta Z_3^b$)
\begin{equation}
\begin{split}
 \delta Z_{3}^a =  
   \frac{\alpha_s}{4\pi} \left[\frac{5}{3} C_A - \frac{4}{3} \, tr \,
    n_f \right] \frac{1}{\epsilon} \, , \qquad
 \delta Z_{3}^b =  
  - \frac{\alpha_s}{4\pi} \left[ \frac{2}{3} \, tr \, n_f + \frac{2}{3} C_A \right] \frac{1}{\epsilon}
  \, ,
\end{split}
\end{equation}
where the first expression stems from gluon, quark and ghost contributions in the corresponding gluon self-energy, while the second
one is due to gluino and squark loops. Note that the second
term corresponds to the UV-singular part of $\delta Z_3^b$ in  (\ref{deltaZ3aonshell}).  
The diagonal renormalization constants for the squark fields are
contained in (\ref{squarkrenminimal}).

The counterterm contribution to the vertex due to the renormalization
of the fields and of $g_{s,G}$ can be written as 
($Z_{g_s,G} = 1 + \delta Z_{g_s,G}^a + \delta Z_{g_s,G}^b$)
\begin{equation}
 \delta \tilde{V}^{ct}_{kk} = 
\left(  \delta \tilde{Z}^a +
\delta \tilde{Z}^b + \frac{1}{2} \, \delta Z^a_3 + \frac{1}{2}
\, \delta Z^b_3  + \delta Z^a_{g_s,G} + \delta Z^b_{g_s,G}
\right) \, \tilde{V}_{kk}^{tree} \, .
\end{equation}
Requiring $\delta \tilde{V}_{kk}^{irred}+\delta \tilde{V}_{kk}^{ct}$
to be finite, fixes 
$\delta Z_{g_s,G}^a$ and
$\delta Z_{g_s,G}^b$
in the $\overline{\rm MS}$ scheme to be
\begin{equation}\begin{split}
  \delta Z_{g_s,G}^{a} =&  \frac{\alpha_s}{4\pi} \left[ \frac{2}{3} \,
    tr\, n_f - \frac{11}{6} C_A \right] \frac{1}{\epsilon} \,,  \\
  \delta Z_{g_s,G}^{b} =&\frac{\alpha_s}{4\pi} 
\left[ \frac{1}{3} \, tr \, n_f + \frac{C_A}{3} \right]
\frac{1}{\epsilon} \,  .
\end{split}\end{equation}

\end{appendix}

\end{document}